\documentclass[useAMS,usenatbib]{mn2e}

\usepackage[dvips]{graphicx}

\title[Rotational Periods in Praesepe]
{Stellar rotational periods in the planet hosting open cluster Praesepe}
\author[G\'eza Kov\'acs et al.]
{G\'eza Kov\'acs$^{1,2}$\thanks{E-mail: kovacs@konkoly.hu}, 
Joel D. Hartman$^{3}$, 
G\'asp\'ar \'A. Bakos$^{3,4}$, 
Samuel N. Quinn$^{6,7}$,\newauthor
Kaloyan Penev$^{3}$,
David W. Latham$^{5}$,
Waqas Bhatti$^{3}$, 
Zolt\'an Csubry$^{3}$ and\newauthor
Miguel de Val-Borro$^{3}$\\
\\
$^{1}$Konkoly Observatory, Budapest, Konkoly Thege M. u. 15-17, Hungary\\
$^{2}$Department of Physics \& Astrophysics, University of North Dakota, 
101 Cornell Street,  Grand Forks, ND 58202, USA \\
$^{3}$Department of Astrophysical Sciences, Princeton University, 
Princeton, NJ 08544, USA \\
$^{4}$Sloan Fellow\\
$^{5}$Harvard-Smithsonian Center for Astrophysics, 60 Garden Street, 
Cambridge, MA 02138, USA\\ 
$^{6}$Department of Physics \& Astronomy, Georgia State University, 
25 Park Place NE Suite 605, Atlanta, GA 30303\\
$^{7}${NSF Graduate Research Fellow}}

\begin{document}

\date{Submitted/Accepted April 3, 2014/May 9, 2014}

\pagerange{\pageref{firstpage}--\pageref{lastpage}} \pubyear{2014}

\maketitle

\label{firstpage}

\begin{abstract}
By using the dense coverage of the extrasolar planet survey project
HATNet, we Fourier analyze $381$ high-probability members of the
nearby open cluster Praesepe (Beehive/M44/NGC~2632). In addition to
the detection of $10$ variables (of $\delta$~Scuti and other types),
we identify $180$ rotational variables (including the two known planet
hosts). This sample increases the number of known rotational variables
in this cluster for spectral classes earlier than M by more than a
factor of three. These stars closely follow a color/magnitude -- period
relation from early F to late K stars.  We approximate this relation
by polynomials for an easier reference to the rotational
characteristics in different colors. The total (peak-to-peak)
amplitudes of the large majority ($94$\%) of these variables span the
range of $0.005$ to $0.04$~mag. The periods cover a range from $2.5$
to $15$~days.  These data strongly confirm that Praesepe and the
Hyades have the same gyrochronological ages. Regarding the two planet
hosts, Pr0211 (the one with the shorter orbital period) has a
rotational period that is $\sim2$~days shorter than the one expected
from the main rotational pattern in this cluster. This, together 
with other examples discussed in the paper, may hint that star-planet 
interaction via tidal dissipation can be significant in some cases in 
the rotational evolution of stars hosting Hot Jupiters.
\end{abstract}



%
\section{Introduction}
The well-known significance of open clusters in studying stellar evolution and 
various aspects of galactic structure and cosmic distance calibration has been 
highlighted recently by the analysis of the photometric databases collected by 
both target-oriented projects, such as the MONITOR project 
\citep[e.g.,][]{irwin2009} and by the ground- and space-based surveys of 
transiting extrasolar planets (TEPs). It is important to note that before these 
surveys and special projects, the data on rotational variables (i.e., on spotted 
stars with measurable photometric variability) were fairly scarce. For example, 
the first few variables of this type were discovered in the otherwise well-known 
nearby cluster Praesepe only in 2007 \citep{scholz2007}. The TEP surveys are very 
powerful means to discover variables, since they stare on the same large area of 
the sky for a period of three to six months (or longer) and gather good quality 
time series on $10^4$--$10^5$ stars in their large field of views. As a 
`by-product' of these surveys, several open clusters were caught and analyzed in 
searching for TEPs and other variables, including spotted stars 
(\cite{pepper2008}, [Praesepe]; \cite{collier2009}, [Coma Berenices]; 
\cite{hartman2010}, [Pleiades]; \cite{delorme2011}, [Hyades, Praesepe]; 
\cite{cargile2013} [Blanco 1]). These latter types of variables play a crucial 
role in studying stellar rotation and its dependence on cluster age and stellar 
type. For a very recent review of this field we refer to \cite{bouvier2013}. 

Earlier studies on stellar rotation via photometric variability were limited 
due to the sporadic nature of the observational campaigns and their focusing 
mostly on individual targets. Furthermore, main sequence stellar rotational 
periods are in the range of few to $10$--$20$~days, which is not a comfortable 
range in following up low-amplitude variables with sparsely sampled 
ground-based observations. In addition, spot evolution and differential 
rotation lead to non-stationary frequency spectra, that may result in an 
inaccurate determination of the rotational periods, and in general, 
misinterpretation of the data. This is why previous works (before the era of 
TEP searches) were mostly limited to rotation effects producing signals above 
$\sim 0.01$~mag \citep[e.g.,][]{messina2002}. Even though most of the 
wide-field ground-based surveys are carried out from a single site,
and thus have daily gaps in their observations, they are still capable
of detecting rotation signals with amplitudes in the {\em sub-mmag}
regime, thereby unveiling the rotation properties of stars with spot
sizes similar to those on the Sun. This sensitivity results from the
dense sampling, long time coverage, and relatively high photometric
precision per observation achieved by these surveys (typical values
are: cadences of a few minutes, $10^3$--$10^4$ total data points per
star, and $0.01$--$0.02$\,mag accuracy per data point).
This dramatically increases the number of stars, from spectral types M
to A, for which stellar rotational periods may be measured, and allows a
deep analysis of angular momentum loss and stellar evolution.

From the point of view of extrasolar planets, open clusters are also
important in addressing questions on the role of stellar environment
in the early phases of planetary system evolution. Although
considerable efforts have been made over the past to find 
planets in clusters using both spectroscopic (i.e., radial velocity)
and photometric (i.e., transit) methods, these searches did not bear
fruits until fairly recently. As of this paper, we know $11$ planets
in six open clusters. We have three long period ($P>600$~days) radial
velocity planets in the Hyades, NGC~2423 and NGC~4349
(\citealt{lovis2007, sato2007}). The other eight systems have shorter
periods. There are two radial velocity planets in Praesepe ($P=4.43$
and $2.15$~days, \citealt{quinn2012}), one in the Hyades ($P=6.09$~days,
\citealt{quinn2014}), three in M67 ($P=6.96$, $5.12$ and
$121.71$~days, \citealt{brucalassi2014}) and two TEPs from the Kepler
satellite in NGC~6811 ($P=15.73$ and $17.82$~days,
\citealt{meibom2013}).  Although this sample is still small, the
number of planets hosted by cluster stars seems to be in agreement
with those derived from field stars (this is true both for Hot
Jupiters and for (sub)Neptunes -- see \citealt{quinn2014} and
\citealt{meibom2013}, respectively).

Since the ages and the chemical compositions of the open clusters above 
are much more accurately known, than, in general, those of the field stars, 
with the five Hot Jupiters (HJs) residing in Praesepe, Hyades and M67, 
one can start addressing several questions (such as planet formation, 
migration and star-planet interaction) in a more efficient way. In particular, 
because of the usually well-defined rotational pattern of cluster stars, 
one may ask if tidal interaction between the planet and the star can lead to 
the spin-up of the host star (\citealt{brown2011, penev2012, zhang2014}). 
The difference between the rotational periods of single and planet host stars 
yields an important constraint on the tidal dissipation factors 
$Q_{\rm star}$ and $Q_{\rm planet}$. There are only bulk estimates on these 
parameters, although they are crucial in any tide-related problems (e.g., 
they scale the circularization and in-spiral times for close-in planets). 

Using photometric time series observations from the
HATNet\footnote{The Hungarian-made Automated Telescope Network
  (\citealt{bakos2004}) consists of 6 wide field of view, small-aperture
  autonomous telescopes located in Hawaii (Mauna Kea) and Arizona
  (Fred Lawrence Whipple Observatory). The prime purpose of the
  project is to search for extrasolar planets via transit 
  technique.} project, here we derive rotational periods for $180$ K
to F stars in Praesepe (M44/NGC~2632/Beehive). We use the tight
period--color (also period--luminosity, or period--mass) relation
spanned by these data to investigate the possibility of orbital
momentum transfer in the two systems containing HJs. The results are
compared with the Hyades, another planet hosting cluster, similar to
Praesepe.

%
\section{Frequency analysis of the cluster members}

%
\subsection{The HATnet data coverage}
In selecting cluster members we rely on the membership analysis of
\cite{kraus2007}, based on various archival color, magnitude and
proper motion data. In their Table~3 there are $1130$ stars listed as
candidate cluster members, an overwhelming majority of which have
large membership probabilities. Here we match all of them with the
HATNet database and find that there are altogether $408$ objects that
match within $0.22"$ of a HAT target (based on the 2MASS catalog). 
The remaining stars in the membership catalog yield much larger matching 
distances (greater than $9"$). This is because they have faint magnitudes 
and, as a result, they are absent from the HATNet database. Therefore, 
we have many inaccurate object identifications among the fainter stars 
with associated bad matching distances. This results in a poor sampling 
of the M dwarf regime, and, consequently, no coverage of the cooler M 
stars, where magnetic braking is not yet effective and the rotational 
periods have significant scatter due to the different initial conditions 
with which stars started their lives in the interstellar cloud. Although 
we have poor sampling in the M star regime, the HATNet survey is able to 
reach fast rotators at the blue end, which is also important, since this 
part of the diagram is still poorly known, largely due to the lower 
amplitudes of these variables. From the $408$ targets we omit $27$ which 
have a small number of data points\footnote{We set the limit to $1000$ 
data points, mainly because when applying systematics filtering (TFA, see 
Sect.~2.2 and \citealt{kovacs2005}) we use a large number of templates; 
stars have different number of data points depending on their brightnesses 
and positions on the sky.} and we end up with a sample of $381$ stars as the 
basic set for the frequency analysis.\footnote{The light curves of these 
stars are deposited at the Strasbourg astronomical Data Center: 
{\bf http://cdsweb.u-strasbg.fr/}}.

%
\begin{figure}
 \vspace{0pt}
 \includegraphics[angle=-90,width=90mm]{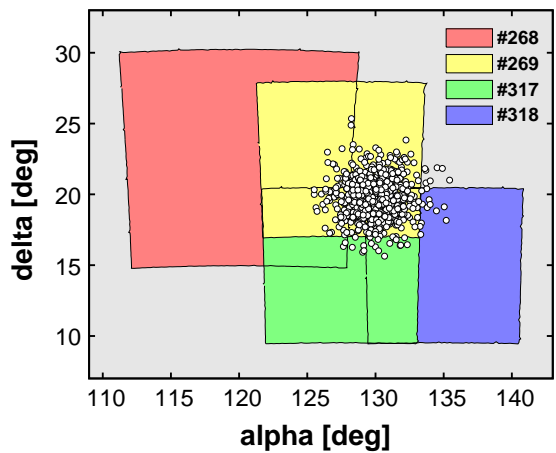}
 \caption{Coverage of Praesepe by the HATNet fields. The $1130$ candidate 
          members given by \protect\cite{kraus2007} are shown as 
	  open circles.}
\label{hat-fields}
\end{figure}

Praesepe has a very fortunate position in respect of the distribution of the 
HATNet fields. The cluster sits in the overlap area of four fields 
(see Fig.~\ref{hat-fields}), yielding long-timebase coverage with high number 
of data points for most of the stars. Although there are four fields, their 
relative positions are such that cluster members may have only up to three-fold 
coverage. Further details of the field-wise data distribution are given in 
Table~\ref{data-journal}. The longest and most abundant coverage comes from 
the overlaps of fields \#269, $317$ and $318$ with the total time span and 
datapoint number of 917~days and 14400, respectively. The integration 
time is $3$~min for all fields, except for \#318, for which we used $5$~min 
integrations. With the read-outs and data storage, the sampling times are 
$\sim 20$\% longer than these integration times. All observations were taken 
through the Sloan {\em r} filter. 

%
\begin{table}
 \centering
 \begin{minipage}{140mm}
  \caption{HATNet fields covering Praesepe.}
  \label{data-journal}
  \begin{tabular}{cccrrrr}
  \hline
   Field  & JD$_{\rm first}$  & JD$_{\rm last}$  & T [days] & N$_{\rm dp}$ 
          & N$_{\rm clus}^{all}$ & N$_{\rm clus}^{act}$\\
 \hline
 268  & 55873 & 56070 & 197  & 7750 &   29 &   2\\
 269  & 55674 & 55728 &  54  & 2750 &  403 & 374\\
 317  & 55507 & 55672 & 165  & 8400 &  337 & 315\\
 318  & 54811 & 54964 & 153  & 2900 &  278 & 241\\
\hline
\end{tabular}
\end{minipage}
\begin{flushleft}
\underline{Notes:} 
JD$_{\rm first}$, JD$_{\rm last}$ (minus $2400000$), T (total time span) and 
N$_{\rm dp}$ (number of data points) may change from object to object. The 
values listed are average values. N$_{\rm clus}^{all}$: all overlaps with the 
cluster member list of \cite{kraus2007}; N$_{\rm clus}^{act}$: actual 
(N$_{\rm dp} > 1000$) overlaps with the list mentioned. Because the same 
object may appear in up to three fields, the total number of HATNet objects 
with cluster-membership and available light curve is only $381$. 
\end{flushleft}
\end{table}

\begin{figure}
 \vspace{0pt}
 \includegraphics[angle=-90,width=85mm]{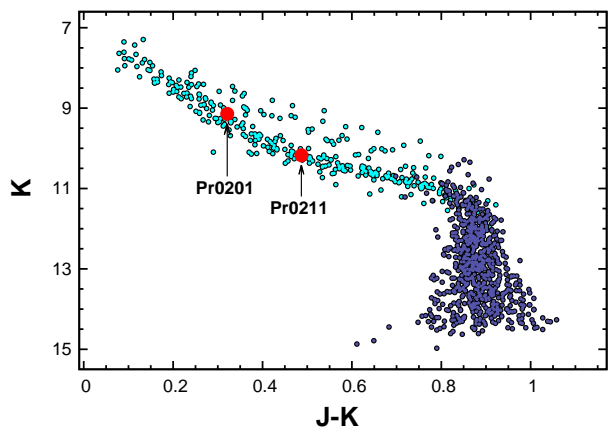}
 \caption{Color-magnitude diagram of Praesepe with the two planet host 
          stars highlighted by larger filled circles. The $1130$ cluster 
	  candidates as given by \protect\cite{kraus2007} are plotted in 
	  the 2MASS color system. The $408$ matching objects with the 
	  HATNet database are shown by light-shaded points.}
\label{j-k-k-sphere}
\end{figure}

The distribution of the cluster member candidates of \cite{kraus2007} on the 
color-magnitude diagram (CMD) is shown in Fig.~\ref{j-k-k-sphere}. The 
lighter/blue shade/color denotes the $408$ HATNet matches, including the two 
radial velocity (RV) planet hosts. This sample extends from early F to late 
K stars and follows a fairly well-defined isochrone corresponding to [Fe/H]$=0.11$ 
and age of $590$~Myr (\citealt{khalaj2013}, see however \citealt{gaspar2009} 
for a possible higher age of $757$~Myr). A particular property of the CMD is 
the distinct high-luminosity branch between F- and early M-type stars. The 
common explanation of this branch is that these stars are either blended by 
some neighbors in the denser inner part of the cluster, or they are binaries 
with luminous secondary components (see, e.g., \citealt{khalaj2013}, 
\citealt{wang2014}). However, we note that an alternative explanation, 
involving an earlier mixing with another stellar population of different age 
may also be viable \citep{franciosini2003}. It is also seen that the two 
planet hosts fit well to the densely-populated main sequence with no apparent 
second luminous components. 
 
%

%
\subsection{Pre-analysis data conditioning} 
Since our time series preparation methods were developed in the context 
of transiting extrasolar planet search, we think it is useful to briefly 
describe the main aspects of our approach. For additional details on the 
two methods summarized below we refer to \cite{kovacs2005} and 
\cite{bakos2010}. 

We use External Parameter Decorrelation (EPD) and Trend Filtering Algorithm 
(TFA) to clean up the data from systematics before the frequency analysis. 
These are crucial steps, since we usually deal with small-amplitude signals, 
that are either completely buried in the systematics or seriously crippled 
by them. 

The idea of the EPD filtering is based on the observation that certain
external parameters \{$p_{\rm i}$\} of the stellar image (such as PSF width,
elongation, chip position, hour angle, color, etc.) correlate with
$\Delta F$, which is the residual stellar flux after subtracting off 
the median of the signal. The functional form of $\Delta F(p_1, p_2,
...)$ is determined numerically through a fit of \{$p_{\rm i}$\} to $\Delta
F$ by using all individual flux measurements for each selected star. 
The correlation is determined for each star separately. Therefore, while 
the functional form of the correlation is the same, the regression 
coefficients will differ for each individual star. Once the correction 
function is found, it is employed on each star on each flux value individually 
by $F_{\rm corr}(t)=F_{\rm in}(t)-\Delta F(p_1(t), p_2(t), ...)$, where 
$F_{\rm in}(t)$ is the incoming flux and \{$p_{\rm i}(t)$\} are the 
time-dependent external parameters. We note that although this step contains 
only the subtraction of a pre-determined function evaluated at the temporal 
values of the EPD parameters and stellar flux measurements, the regression 
is obtained by a fit to the target time series, and, as a result, the 
underlying signal might also be affected by this process. Our experience 
shows that this effect is usually small, and can be corrected at a later 
step with full time series model fitting (see \citealt{bakos2010}). 

TFA has a different philosophy in finding systematics in photometric time 
series. Akin in part to the traditional ensemble photometry, it is assumed 
that systematics exhibit themselves in many objects in a similar manner. 
In the case of standard ensemble photometry this behavior is simply utilized 
by calculating the average flux of a large number of neighboring stars and 
dividing the flux of the target by this average. However, in the case of TFA 
we assume that not all stars have the same systematics but for any given 
target we can find a time series \{$Z(t_{\rm i})$\} that is optimum in terms 
of systematics (i.e., it contains most of them, characteristic for the target 
signal). The optimization is achieved by using the time series of $m$ number 
of comparison stars (or templates) \{$C_{\rm j}(t_{\rm i})$\} and searching 
for the optimum filter \{$U(t_{\rm i})$\} in the form of the linear combination 
of these template time series: 
$U(t_{\rm i})=\sum_{\rm j=1}^{\rm m}a_{\rm j}C_{\rm j}(t_{\rm i})$. The 
regression coefficients \{$a_{\rm j}$\} can be found, e.g., by fitting 
\{$U(t_{\rm i})$\} to \{$Z(t_{\rm i})$\} by least squares. The filtered time 
series is obtained by subtracting \{$U(t_{\rm i})$\} from \{$Z(t_{\rm i})$\}. 
In this process we assume that the target time series is dominated by 
systematics - the non-reconstructive application of TFA. This assumption can 
be lifted once a signal is found, and a complete model is fitted that includes 
also the signal -- the reconstructive application of TFA).

%
\subsection{Frequency analysis}
We employ the method of standard Discrete Fourier Transformation (DFT, 
\citealt{deeming1975}) to search for significant close-to-sinusoidal signals 
in the above dataset of $381$ light curves (LCs). The analysis is performed 
both on the basic product of the HATNet pipeline, comprising the time series 
after EPD and on the TFA-filtered time series of the EPD data. When analyzing 
the EPD data, we merge the LCs from the individual fields by shifting their 
averages to the same level. For the TFA-filtered data we perform 
non-reconstructive TFA filtering for each field separately, and then we merge 
the so-obtained LCs in the same way as we do for the EPD data. The analyses of 
the two types of datasets are complementary. The EPD data still contain various 
systematics, thereby weakening our ability to detect faint signals. On the other 
hand, application of TFA-filtering might lead to losing some of the variables 
due to over filtering, especially if they are on the longer period side and if 
there are other stars in the field with similar periods. Although our earlier 
works on different datasets (\citealt{szulagyi2009, dekany2009}) showed that 
the chance of this to happen is reasonably low, incorporating the original (EPD) 
data may further reduce the likelihood of this event. Furthermore, because 
several/many rotational variables in a cluster exhibit very similar periods, 
to minimize the effect of `squeezing' the amplitudes when TFA is employed, we 
iteratively deselect suspected variables from the TFA templates. 

In the variability search we test narrow-, medium- and high-frequency bands 
for the EPD and the TFA-filtered data. For the latter, we use $600$ templates 
selected as mentioned above. For the frequency bands we choose the ranges of 
$[0,0.5]$d$^{-1}$, $[0,10]$d$^{-1}$ and $[10,50]$d$^{-1}$. Both the frequency 
spectra and the folded LCs are examined. At the end of this process we conclude 
with $180$ rotational and $10$, mostly $\delta$~Scuti-type variables. The latter 
ones are summarized in the Appendix, whereas the rotational variables (including 
the selection procedure) are discussed in the subsequent sections.

%
\section{Rotational variables}
One of the distinct properties of main sequence cluster stars is that their 
rotational rates quickly (within a few hundred Myrs) settle down to a fairly 
tight sequence, usually parameterized by their mass, or by some other, 
observationally more easily accessible parameters, such as their colors 
\citep{barnes2003}. Therefore, unlike in the case of rotating (i.e., spotted) 
variables in the field, they are much easier to recognize, especially if 
the cluster is rich, older than $\sim100$~Myr and the underlying photometric 
data are accurate and extended enough to detect (quasi)periodic variation 
down to the mmag level. 

%
\begin{figure}
 \vspace{0pt}
 \includegraphics[angle=-90,width=85mm]{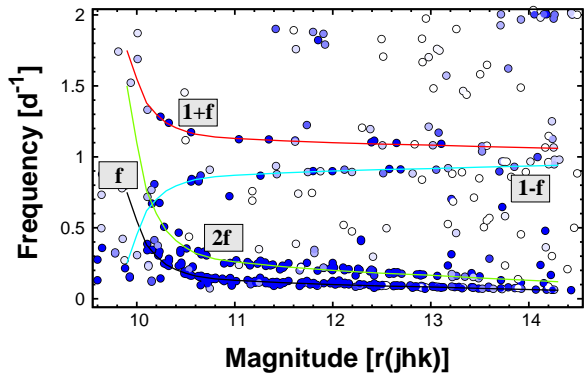}
 \caption{Frequency versus approximate Sloan r(jhk) magnitudes for 
          our Praesepe sample. All frequencies found as the highest 
	  peaks in the frequency spectra of the data and in their residuals 
	  after the first pre-whitening are plotted (i.e., two points 
	  for each star). The points are color/shade-coded from high 
	  signal-to-noise ratio (SNR$>8$, dark) to the low one (SNR$<5$, 
	  white). The analysis was made in the $[0,2]$~d$^{-1}$ frequency 
	  band. The derived rotation line (see text) is used to show the 
	  various frequency patterns due to aliases and confusion with 
	  harmonic components.}
\label{rmag-freq-all}
\end{figure}

As an example of the striking appearance of rotational variables, in 
Fig.~\ref{rmag-freq-all} we show all derived frequencies in the 
r(jhk)--frequency plane. The analysis covers the $[0,2]$d$^{-1}$ range 
and includes a single pre-whitening step for each star (i.e., we plot 
two frequency values for each of the $381$ stars: the peak frequencies 
obtained from the original and the pre-whitened data). The color coding for the 
signal-to-noise ratio (SNR) helps to recognize more easily the various 
ridges related to the main rotational sequence. The magnitudes in the 
Sloan {\em r} band are approximated by the following combination of the 
2MASS colors (based on our linear regression to $410$ stars of 
\citealt{stetson2000} with colors covering the range relevant for 
Praesepe)\footnote{Because of the several non-matching objects with the 
APASS database ({\bf http://www.aavso.org/apass})} for Fig.~\ref{rmag-freq-all}, 
we decided to use exclusively these approximate Sloan {\em r} magnitudes, 
instead of mixing the two sets. In all other cases either the APASS or the 
2MASS magnitudes are employed. We avoid using HATNet instrumental `r' 
magnitudes, since the HATNet magnitudes are often unreliable, due to the 
limited spatial resolution of the images. 

\begin{equation}
{\rm r(jhk)} = 0.6975 + 2.9782J - 0.8809H - 1.1230K \hskip 2mm .
\end{equation}

\noindent
The rotational sequence (labelled by the letter `f') shows up very clearly, 
as does the one related to the 1st harmonic (sequence `2f'). Other 
components, due to $1$~d$^{-1}$ aliasing (sequences `$1\pm f$') show up too, 
albeit with a sparser population, indicating that in most cases aliasing 
is not an issue in the available dataset.

%
\begin{figure}
 \vspace{0pt}
 \includegraphics[angle=-90,width=85mm]{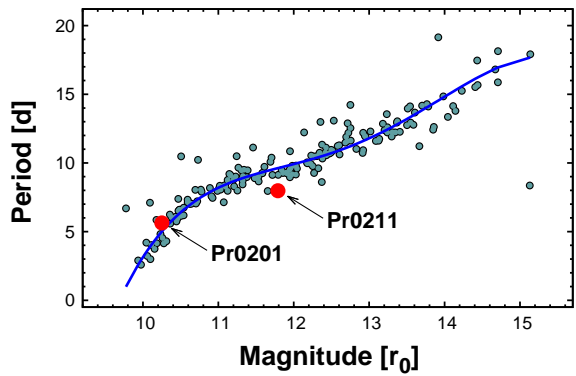}
 \caption{Dereddened Sloan {\em r} magnitudes versus period for our 
          rotational variable star sample of $180$ stars in Praesepe (see 
	  Table~\ref{data-rot-var}). The continuous line is a 4th-order 
	  polynomial fit to the data (see Table~\ref{polfit-rot-var}). 
	  The two RV planet hosts of \protect\cite{quinn2012} are shown 
	  by the larger filled circles.}
\label{rmag-period-2rvp}
\end{figure}

After the identification of the main rotational ridge, we go through 
the multistep variability check as described in Sect.~2.3 and construct 
the final rotational ridge by correcting the peak frequencies, if they 
belong to some of the secondary ridges mentioned above. Here we extend 
the correction due to harmonic components up to the 4th harmonic 
-- because of the degeneracy between the observed and the rotational periods 
we may need to consider also these, more involved cases.\footnote{A simple 
example of this degeneracy is when two spots of equal size are at the same 
latitude but at opposite longitudes. The resulting photometric signal will 
have a period of half that of the rotational period.} 
In constructing the ridge, we iteratively develop the average ridge line, 
which is a 4th-order polynomial fit to the data. In deciding whether 
a frequency value should or should not be corrected for harmonics or 
aliases, we consider the difference between this frequency point 
and the one corresponding to the ridge value (at the given {\em r} 
magnitude). The consecutive steps in the frequency correction and 
polynomial fitting constitutes the iterative process.   

%
%
\setcounter{table}{1}
\begin{table*}                                                    
 \centering                                                       
 \begin{minipage}{200mm}                                          
  \caption{HATNet rotational variables in the Praesepe.}          
  \label{data-rot-var}                                            
  \scalebox{0.90}{                                                
  \begin{tabular}{cclccccccccr}                                   
  \hline                                                          
HAT ID/type & 2MASS ID & $\nu$      &  1/T       &   B   &   V       &   r      &    K       & J$-$H      & J$-$K &   A       &  SNR\\
     &          & [d$^{-1}$] & [d$^{-1}$] & [mag] & [mag]     & [mag]    & [mag]      &  [mag]     & [mag] & [mag]     & \\
\hline                                                            
HAT-269-0000500 b  & 08401762+194715 &  0.149427 &  0.0011 & 10.551 &  9.941 &  9.837 &  8.583 &  0.267 &  0.303 &  0.0044 & 10.5\\         
HAT-269-0000588 b  & 08402231+200624 &  0.386161 &  0.0011 & 10.554 & 10.071 & 10.036 &  8.852 &  0.213 &  0.285 &  0.0036 &  7.6\\         
HAT-269-0000592 b  & 08411002+193032 &  0.345078 &  0.0011 & 10.613 & 10.080 &  9.999 &  8.908 &  0.243 &  0.269 &  0.0045 &  9.4\\         
HAT-269-0000667 b  & 08395234+191845 &  0.309961 &  0.0011 & 10.742 & 10.217 & 10.119 &  9.008 &  0.192 &  0.275 &  0.0042 &  8.3\\         
HAT-269-0000673 b  & 08400062+194823 &  0.333975 &  0.0011 & 10.824 & 10.271 & 10.161 &  9.079 &  0.184 &  0.251 &  0.0072 &  6.5\\         
HAT-269-0000610 d  & 08393042+200408 &  0.140742 &  0.0011 & 10.892 & 10.260 & 10.155 &  8.807 &  0.301 &  0.333 &  0.0054 &  5.9\\         
HAT-269-0000701 d  & 08405252+192859 &  0.248768 &  0.0011 & 10.839 & 10.267 & 10.173 &  9.049 &  0.183 &  0.288 &  0.0047 &  5.7\\         
HAT-317-0000737 b  & 08413154+183002 &  0.240999 &  0.0011 & 10.886 & 10.384 & 10.246 &  9.105 &  0.221 &  0.259 &  0.0075 & 12.8\\         
HAT-317-0000748 c  & 08391096+181033 &  0.238181 &  0.0011 & 10.748 & 10.218 & 10.107 &  8.994 &  0.237 &  0.313 &  0.0031 &  6.3\\         
HAT-270-0000544 c  & 08450422+202127 &  0.267231 &  0.0045 & 10.865 & 10.296 & 10.216 &  9.130 &  0.218 &  0.259 &  0.0060 &  6.8\\         
HAT-269-0000768 b  & 08382429+200621 &  0.232577 &  0.0011 & 11.085 & 10.521 & 10.373 &  9.182 &  0.221 &  0.281 &  0.0161 & 14.7\\         
HAT-318-0000612 c  & 08480173+184037 &  0.170922$\times2$ &  0.0011 & 11.043 & 10.440 & 10.241 &  8.806 &  0.321 &  0.422 &  0.0101 &  9.3\\
HAT-269-0000761 c  & 08392498+192733 &  0.241306 &  0.0011 & 11.066 & 10.461 & 10.339 &  8.997 &  0.269 &  0.360 &  0.0073 &  8.0\\         
HAT-269-0000805 c  & 08414382+201336 &  0.177573 &  0.0011 & 10.988 & 10.431 & 10.313 &  9.142 &  0.236 &  0.321 &  0.0066 &  8.2\\         
HAT-318-0000683 d  & 08471411+162347 &  0.174265 &  0.0012 & 11.200 & 10.608 & 10.506 &  9.294 &  0.222 &  0.267 &  0.0155 & 12.8\\         
HAT-270-0000640 b  & 08452794+213912 &  0.207376 &  0.0185 & 11.046 & 10.435 & 10.294 &  9.257 &  0.227 &  0.294 &  0.0082 &  5.0\\         
HAT-269-0000861 b  & 08394575+192201 &  0.164439 &  0.0011 & 11.257 & 10.640 & 10.497 &  9.259 &  0.208 &  0.306 &  0.0089 &  8.9\\         
HAT-269-0000834 b  & 08391217+190656 &  0.179223 &  0.0011 & 11.129 & 10.568 & 10.424 &  9.258 &  0.242 &  0.297 &  0.0040 &  6.6\\         
HAT-270-0000678 b  & 08464732+193841 &  0.162909 &  0.0011 & 11.303 & 10.701 & 10.566 &  9.342 &  0.199 &  0.279 &  0.0070 &  9.2\\         
HAT-269-0000794 b  & 08443703+194239 &  0.220523 &  0.0011 & 11.037 & 10.477 & 10.321 &  9.054 &  0.322 &  0.366 &  0.0081 &  9.8\\         
HAT-269-0000877 d  & 08374660+192618 &  0.168159 &  0.0011 & 11.243 & 10.635 & 10.521 &  9.276 &  0.261 &  0.315 &  0.0095 &  7.1\\         
HAT-269-0000850 b  & 08404189+191325 &  0.160327$\times2$ &  0.0011 & 11.205 & 10.567 & 10.419 &  9.063 &  0.311 &  0.396 &  0.0173 &  3.8\\
HAT-269-0000905 b  & 08402271+192753 &  0.153421 &  0.0011 & 11.303 & 10.673 & 10.533 &  9.336 &  0.239 &  0.304 &  0.0096 & 12.8\\         
HAT-269-0000871 d  & 08424525+185136 &  0.095443 &  0.0011 & 11.449 & 10.766 & 10.569 &  9.168 &  0.299 &  0.362 &  0.0047 & 11.7\\         
HAT-269-0000913 d  & 08432019+194608 &  0.161667 &  0.0011 & 11.415 & 10.756 & 10.602 &  9.359 &  0.243 &  0.300 &  0.0103 & 10.9\\         
HAT-269-0000885 b  & 08410961+195118 &  0.165487$\times2$ &  0.0011 & 11.361 & 10.631 & 10.440 &  8.939 &  0.342 &  0.467 &  0.0114 & 15.6\\
HAT-317-0000999 b  & 08373307+183915 &  0.135465 &  0.0011 & 11.249 & 10.668 & 10.531 &  9.283 &  0.274 &  0.338 &  0.0069 & 12.0\\         
HAT-269-0000946 d  & 08412584+195636 &  0.151615 &  0.0011 & 11.342 & 10.687 & 10.546 &  9.329 &  0.263 &  0.330 &  0.0087 &  9.9\\         
HAT-269-0000936 b  & 08393553+185236 &  0.151867 &  0.0011 & 11.282 & 10.695 & 10.550 &  9.329 &  0.276 &  0.328 &  0.0176 & 14.4\\         
HAT-270-0000714 c  & 08451310+194112 &  0.136573 &  0.0011 & 11.602 & 10.848 & 10.655 &  9.158 &  0.294 &  0.403 &  0.0067 &  8.7\\         
HAT-317-0001078 b  & 08362782+175453 &  0.138810 &  0.0045 & 11.374 & 10.810 & 10.659 &  9.405 &  0.235 &  0.315 &  0.0083 &  5.8\\         
HAT-317-0001111 b  & 08355455+180857 &  0.142247 &  0.0045 & 11.427 & 10.854 & 10.705 &  9.445 &  0.239 &  0.312 &  0.0059 &  9.0\\         
HAT-269-0000993 d  & 08403992+194009 &  0.140088 &  0.0011 & 11.580 & 10.859 & 10.656 &  9.189 &  0.299 &  0.421 &  0.0113 & 13.5\\         
HAT-269-0001025 b  & 08305546+193319 &  0.148258 &  0.0045 & 11.384 & 10.736 & 10.603 &  9.369 &  0.262 &  0.356 &  0.0119 & 11.2\\         
HAT-269-0001089 c  & 08412869+194448 &  0.141959 &  0.0011 & 11.605 & 10.933 & 10.765 &  9.469 &  0.263 &  0.343 &  0.0057 & 10.9\\         
HAT-269-0001128 d  & 08415587+194122 &  0.124125 &  0.0011 & 11.718 & 11.006 & 10.830 &  9.544 &  0.242 &  0.325 &  0.0127 & 13.2\\         
HAT-269-0001075 d  & 08404798+193932 &  0.138748 &  0.0011 & 11.757 & 10.991 & 10.782 &  9.254 &  0.353 &  0.439 &  0.0093 & 13.9\\         
HAT-269-0001149 d  & 08404832+195518 &  0.126479 &  0.0011 & 11.707 & 11.012 & 10.840 &  9.507 &  0.276 &  0.357 &  0.0155 & 11.9\\         
HAT-269-0001058 d  & 08381427+192155 &  0.142596 &  0.0011 & 11.605 & 10.897 & 10.685 &  9.188 &  0.375 &  0.464 &  0.0097 & 10.8\\         
HAT-269-0001147 b  & 08423225+192346 &  0.097807 &  0.0011 & 11.656 & 10.954 & 10.793 &  9.458 &  0.300 &  0.378 &  0.0069 &  9.8\\         
HAT-269-0001124 d  & 08400635+191826 &  0.124386 &  0.0011 & 11.761 & 10.980 & 10.757 &  9.228 &  0.380 &  0.473 &  0.0067 & 11.3\\         
HAT-269-0001249 d  & 08345963+210549 &  0.122355 &  0.0185 & 11.874 & 11.114 & 10.950 &  9.684 &  0.251 &  0.328 &  0.0156 &  8.3\\         
HAT-269-0001268 d  & 08402743+191640 &  0.131965 &  0.0011 & 11.822 & 11.148 & 10.971 &  9.655 &  0.314 &  0.357 &  0.0129 & 14.0\\         
HAT-269-0001333 c  & 08415437+191526 &  0.120981$\times2$ &  0.0011 & 12.104 & 11.354 & 11.156 &  9.643 &  0.294 &  0.383 &  0.0347 & 19.1\\
HAT-269-0001377 b  & 08391499+201238 &  0.119970 &  0.0011 & 11.946 & 11.278 & 11.082 &  9.649 &  0.272 &  0.393 &  0.0071 & 14.0\\         
HAT-269-0001340 b  & 08415924+205507 &  0.121889 &  0.0185 & 11.958 & 11.259 & 11.076 &  9.705 &  0.303 &  0.362 &  0.0098 &  3.6\\         
HAT-269-0001410 b  & 08375208+195913 &  0.125548 &  0.0011 & 11.952 & 11.242 & 11.051 &  9.689 &  0.288 &  0.390 &  0.0159 & 10.8\\         
HAT-269-0001400 c  & 08371148+194813 &  0.124051 &  0.0011 & 12.008 & 11.286 & 11.093 &  9.686 &  0.295 &  0.389 &  0.0066 &  7.3\\         
HAT-269-0001405 d  & 08404248+193357 &  0.119527 &  0.0011 & 12.063 & 11.331 & 11.142 &  9.706 &  0.307 &  0.385 &  0.0064 &  8.0\\         
HAT-269-0001431 b  & 08381497+203404 &  0.114133 &  0.0185 & 12.100 & 11.400 & 11.186 &  9.717 &  0.302 &  0.389 &  0.0103 &  9.8\\         
HAT-269-0001352 c  & 08374998+195328 &  0.140085 &  0.0011 & 12.265 & 11.403 & 11.134 &  9.326 &  0.401 &  0.535 &  0.0135 & 12.5\\         
HAT-269-0001402 d  & 08364896+191526 &  0.134236 &  0.0045 & 11.945 & 11.212 & 10.986 &  9.687 &  0.336 &  0.396 &  0.0109 &  7.9\\         
HAT-269-0001517 d  & 08372222+201037 &  0.125524 &  0.0011 & 12.077 & 11.378 & 11.186 &  9.803 &  0.289 &  0.384 &  0.0085 & 11.1\\         
HAT-269-0001467 d  & 08371829+194156 &  0.110117 &  0.0011 & 12.181 & 11.424 & 11.223 &  9.802 &  0.326 &  0.373 &  0.0039 &  7.9\\         
HAT-318-0001287 c  & 08482783+182043 &  0.108759 &  0.0011 & 12.111 & 11.411 & 11.208 &  9.772 &  0.348 &  0.392 &  0.0061 &  8.0\\         
HAT-269-0001515 d  & 08404761+185411 &  0.109594 &  0.0011 & 12.159 & 11.426 & 11.221 &  9.686 &  0.348 &  0.440 &  0.0059 &  8.2\\         
HAT-269-0001544 b  & 08430055+202016 &  0.115043 &  0.0045 & 12.147 & 11.384 & 11.189 &  9.774 &  0.332 &  0.411 &  0.0052 &  9.1\\         
HAT-269-0001510 d  & 08361639+193231 &  0.120615 &  0.0045 & 12.458 & 11.573 & 11.272 &  9.520 &  0.390 &  0.511 &  0.0123 &  7.9\\         
HAT-269-0001490 b  & 08392858+192825 &  0.095355 &  0.0011 & 12.519 & 11.606 & 11.305 &  9.533 &  0.409 &  0.502 &  0.0062 & 11.6\\         
HAT-269-0001570 b  & 08411031+194907 &  0.111032 &  0.0011 & 12.268 & 11.495 & 11.292 &  9.753 &  0.358 &  0.430 &  0.0096 &  9.4\\         
\hline                                                            
\end{tabular}}                                                    
\end{minipage}                                                  
\end{table*}                                                      

\setcounter{table}{1}
\begin{table*}                                                    
 \centering                                                       
 \begin{minipage}{200mm}                                          
  \caption{HATNet rotational variables in the Praesepe. -- Continued}                                             
  \scalebox{0.90}{                                                
  \begin{tabular}{cclccccccccr}                                   
  \hline                                                          
HAT ID/type & 2MASS ID & $\nu$      &  1/T       &   B   &   V       &   r      &    K       & J$-$H      & J$-$K &   A       &  SNR\\
     &          & [d$^{-1}$] & [d$^{-1}$] & [mag] & [mag]     & [mag]    & [mag]      &  [mag]     & [mag] & [mag]     & \\
\hline
HAT-317-0001886 a  & 08403360+184028 &  0.105660 &  0.0011 & 12.335 & 11.596 & 11.368 &  9.832 &  0.306 &  0.417 &  0.0054 & 12.9\\         
HAT-269-0001637 c  & 08403169+195101 &  0.118601 &  0.0011 & 12.402 & 11.609 & 11.384 &  9.911 &  0.309 &  0.383 &  0.0035 &  9.5\\         
HAT-317-0001780 d  & 08402440+182713 &  0.103601 &  0.0011 & 12.368 & 11.584 & 11.328 &  9.599 &  0.406 &  0.503 &  0.0090 &  9.8\\         
HAT-269-0001650 d  & 08403184+201206 &  0.113741 &  0.0011 & 12.201 & 11.472 & 11.265 &  9.831 &  0.356 &  0.428 &  0.0055 &  8.6\\         
HAT-269-0001741 d  & 08364572+200726 &  0.110516 &  0.0045 & 12.380 & 11.613 & 11.389 &  9.938 &  0.305 &  0.402 &  0.0139 &  8.8\\         
HAT-269-0001710 d  & 08424250+190558 &  0.113111 &  0.0011 & 12.359 & 11.596 & 11.376 &  9.881 &  0.324 &  0.421 &  0.0159 & 14.2\\         
HAT-269-0001674 d  & 08372755+193703 &  0.114988 &  0.0011 & 12.416 & 11.616 & 11.385 &  9.806 &  0.361 &  0.446 &  0.0140 & 14.5\\         
HAT-269-0001632 d  & 08402863+201844 &  0.125217 &  0.0011 & 12.432 & 11.604 & 11.315 &  9.462 &  0.457 &  0.586 &  0.0071 &  8.1\\         
HAT-269-0001790 d  & 08380808+202622 &  0.111455 &  0.0185 & 12.432 & 11.684 & 11.441 &  9.932 &  0.346 &  0.426 &  0.0154 &  5.3\\         
HAT-269-0001789 d  & 08403623+213342 &  0.105293 &  0.0185 & 12.423 & 11.629 & 11.402 &  9.974 &  0.365 &  0.412 &  0.0077 &  7.7\\         
HAT-269-0001917 c  & 08413384+195808 &  0.105716 &  0.0011 & 12.503 & 11.678 & 11.451 &  9.928 &  0.309 &  0.464 &  0.0039 &  6.9\\         
HAT-269-0001771 a  & 08395983+193400 &  0.089919 &  0.0011 & 12.880 & 11.886 & 11.523 &  9.484 &  0.476 &  0.622 &  0.0120 & 14.8\\         
HAT-269-0001893 d  & 08431076+193134 &  0.112537 &  0.0011 & 12.583 & 11.769 & 11.551 & 10.007 &  0.365 &  0.432 &  0.0141 & 13.7\\         
HAT-269-0001932 b  & 08375703+191410 &  0.115590 &  0.0011 & 12.632 & 11.829 & 11.597 & 10.038 &  0.350 &  0.432 &  0.0108 & 12.4\\         
HAT-269-0001910 d  & 08431784+203037 &  0.112342 &  0.0185 & 12.610 & 11.785 & 11.544 &  9.776 &  0.427 &  0.542 &  0.0245 &  4.7\\         
HAT-269-0002062 d  & 08400968+193717 &  0.125705 &  0.0011 & 12.808 & 11.973 & 11.717 & 10.133 &  0.316 &  0.423 &  0.0141 &  9.8\\         
HAT-269-0001961 b  & 08392155+204529 &  0.108068 &  0.0185 & 12.702 & 11.902 & 11.645 & 10.026 &  0.381 &  0.450 &  0.0132 &  6.2\\         
HAT-317-0002172 b  & 08405967+182204 &  0.107968 &  0.0011 & 12.906 & 11.997 & 11.646 &  9.683 &  0.481 &  0.578 &  0.0049 &  8.0\\         
HAT-270-0001619 d  & 08490670+194111 &  0.102851 &  0.0011 & 12.730 & 11.907 & 11.673 & 10.068 &  0.360 &  0.465 &  0.0079 &  9.5\\         
HAT-269-0002072 b  & 08351780+193810 &  0.111128 &  0.0045 & 12.631 & 11.822 & 11.595 & 10.012 &  0.380 &  0.485 &  0.0071 &  7.6\\         
HAT-269-0002135 b  & 08400416+194703 &  0.096288 &  0.0011 & 12.899 & 12.030 & 11.768 & 10.004 &  0.384 &  0.503 &  0.0076 & 10.8\\         
HAT-269-0002031 d  & 08424372+193723 &  0.099158 &  0.0011 & 13.114 & 12.109 & 11.754 &  9.800 &  0.456 &  0.571 &  0.0059 &  8.2\\         
HAT-269-0002148 d  & 08380758+195916 &  0.109603 &  0.0011 & 12.969 & 12.061 & 11.780 &  9.898 &  0.448 &  0.559 &  0.0157 & 12.6\\         
HAT-269-0002273 d  & 08374739+190624 &  0.107563 &  0.0011 & 13.172 & 12.273 & 11.960 & 10.202 &  0.414 &  0.464 &  0.0155 & 12.4\\         
HAT-269-0002449 d  & 08390228+191934 &  0.102145 &  0.0011 & 13.306 & 12.366 & 12.043 & 10.259 &  0.358 &  0.471 &  0.0113 &  9.5\\         
HAT-269-0002316 d  & 08421149+191637 &  0.125418 &  0.0011 & 13.060 & 12.143 & 11.852 & 10.173 &  0.418 &  0.487 &  0.0170 & 13.3\\         
HAT-269-0002419 d  & 08424021+190759 &  0.109495 &  0.0011 & 12.976 & 12.111 & 11.856 & 10.186 &  0.391 &  0.499 &  0.0092 & 11.6\\         
HAT-269-0002450 b  & 08374640+193557 &  0.105249 &  0.0011 & 13.252 & 12.305 & 12.007 & 10.244 &  0.384 &  0.481 &  0.0153 & 13.0\\         
HAT-269-0002492 b  & 08405669+194405 &  0.107877 &  0.0011 & 13.095 & 12.198 & 11.920 & 10.213 &  0.404 &  0.506 &  0.0132 & 10.8\\         
HAT-269-0002593 d  & 08414368+195743 &  0.108949 &  0.0011 & 13.264 & 12.333 & 12.042 & 10.258 &  0.398 &  0.507 &  0.0114 & 14.1\\         
HAT-269-0002482 b  & 08433880+221609 &  0.111669 &  0.0185 & 13.222 & 12.323 & 12.012 & 10.251 &  0.442 &  0.494 &  0.0112 &  4.4\\         
HAT-269-0002573 d  & 08362269+191129 &  0.101991 &  0.0045 & 13.351 & 12.418 & 12.092 & 10.257 &  0.417 &  0.506 &  0.0212 & 11.9\\         
HAT-317-0002919 b  & 08441706+184411 &  0.092146 &  0.0011 & 13.165 & 12.310 & 12.013 & 10.262 &  0.416 &  0.504 &  0.0041 &  9.2\\         
HAT-269-0002527 b  & 08403347+193800 &  0.105467 &  0.0011 & 13.094 & 12.198 & 11.902 & 10.169 &  0.444 &  0.537 &  0.0162 & 14.3\\         
HAT-269-0002623 d  & 08433239+194437 &  0.103944 &  0.0011 & 13.246 & 12.309 & 12.021 & 10.218 &  0.435 &  0.537 &  0.0118 & 13.5\\         
HAT-317-0002964 b  & 08393752+181013 &  0.108451 &  0.0011 & 13.128 & 12.269 & 11.978 & 10.237 &  0.442 &  0.526 &  0.0132 & 13.7\\         
HAT-269-0002678 a  & 08384610+203436 &  0.109185 &  0.0185 & 13.282 & 12.407 & 12.090 & 10.313 &  0.426 &  0.512 &  0.0148 &  8.4\\         
HAT-269-0002756 d  & 08392185+195140 &  0.081829 &  0.0011 & 13.487 & 12.517 & 12.200 & 10.369 &  0.424 &  0.505 &  0.0076 &  8.8\\         
HAT-269-0002682 b  & 08400571+190130 &  0.102428$\times2$ &  0.0011 & 13.720 & 12.706 & 12.314 & 10.009 &  0.509 &  0.649 &  0.0466 & 17.9\\
HAT-269-0002812 d  & 08410725+192648 &  0.104281 &  0.0011 & 13.370 & 12.416 & 12.104 & 10.288 &  0.445 &  0.553 &  0.0131 & 11.4\\         
HAT-269-0002804 b  & 08393836+192627 &  0.094945 &  0.0011 & 13.627 & 12.648 & 12.294 & 10.286 &  0.453 &  0.553 &  0.0114 &  9.6\\         
HAT-269-0002837 b  & 08413070+185218 &  0.104063$\times4$ &  0.0011 & 13.649 & 12.727 & 12.388 & 10.154 &  0.463 &  0.615 &  0.0553 & 16.1\\
HAT-269-0002769 a  & 08401571+195454 &  0.100648 &  0.0011 & 13.722 & 12.677 & 12.287 & 10.013 &  0.527 &  0.667 &  0.0132 & 12.3\\         
HAT-269-0003067 d  & 08435467+185336 &  0.095532 &  0.0011 & 13.600 & 12.671 & 12.329 & 10.412 &  0.437 &  0.549 &  0.0153 & 13.3\\         
HAT-269-0087638 d  & 08361410+193717 &  0.100242 &  0.0045 & 13.507 & 12.561 & 12.241 & 10.363 &  0.464 &  0.568 &  0.0077 &  7.2\\         
HAT-269-0003068 b  & 08362830+201342 &  0.099556$\times2$ &  0.0045 & 13.521 & 12.564 & 12.240 & 10.441 &  0.460 &  0.541 &  0.0066 &  5.8\\
HAT-269-0002891 d  & 08401893+201130 &  0.077013 &  0.0011 & 14.104 & 12.912 & 12.417 & 10.038 &  0.562 &  0.691 &  0.0092 & 11.9\\         
HAT-269-0003096 b  & 08365411+184524 &  0.107872 &  0.0045 & 13.626 & 12.688 & 12.332 & 10.469 &  0.461 &  0.534 &  0.0111 & 10.8\\         
HAT-269-0003250 b  & 08394707+194939 &  0.095265 &  0.0011 & 13.724 & 12.724 & 12.368 & 10.435 &  0.449 &  0.571 &  0.0088 &  8.9\\         
HAT-317-0003872 b  & 08373821+182857 &  0.088607 &  0.0011 & 14.147 & 13.082 & 12.665 & 10.427 &  0.419 &  0.590 &  0.0262 & 18.6\\         
HAT-269-0003348 d  & 08322347+205944 &  0.089952 &  0.0185 & 13.872 & 12.861 & 12.490 & 10.470 &  0.450 &  0.573 &  0.0169 &  4.7\\         
HAT-317-0003864 a  & 08364711+183446 &  0.096869 &  0.0045 & 13.689 & 12.769 & 12.442 & 10.586 &  0.439 &  0.528 &  0.0135 & 13.1\\         
HAT-270-0002625 a  & 08501855+192542 &  0.116183 &  0.0011 & 13.680 & 12.745 & 12.434 & 10.526 &  0.432 &  0.563 &  0.0045 &  5.8\\         
HAT-269-0003407 b  & 08340436+203430 &  0.093106 &  0.0185 & 13.843 & 12.769 & 12.405 & 10.442 &  0.470 &  0.596 &  0.0194 &  5.1\\         
HAT-269-0003424 b  & 08280099+195417 &  0.094660 &  0.0045 & 13.731 & 12.734 & 12.423 & 10.490 &  0.470 &  0.580 &  0.0053 &  6.6\\         
HAT-269-0003584 d  & 08354516+193826 &  0.091304 &  0.0045 & 13.955 & 12.925 & 12.559 & 10.539 &  0.443 &  0.580 &  0.0116 &  6.8\\         
HAT-269-0003387 d  & 08444870+201725 &  0.088264 &  0.0011 & 13.912 & 12.834 & 12.438 & 10.508 &  0.502 &  0.572 &  0.0135 & 11.2\\         
HAT-269-0003664 d  & 08325223+195835 &  0.089051 &  0.0045 & 14.138 & 13.050 & 12.645 & 10.560 &  0.462 &  0.585 &  0.0206 & 10.4\\         
HAT-269-0003588 d  & 08395998+193440 &  0.094691 &  0.0011 & 13.861 & 12.855 & 12.495 & 10.552 &  0.484 &  0.582 &  0.0198 & 13.4\\         
HAT-270-0002790 b  & 08465012+210112 &  0.076449 &  0.0189 & 14.128 & 13.013 & 12.593 & 10.544 &  0.475 &  0.597 &  0.0098 &  7.4\\         
\hline                                                            
\end{tabular}}                                                    
\end{minipage}                                                  
\end{table*}                                                      

\setcounter{table}{1}
\begin{table*}                                                    
 \centering                                                       
 \begin{minipage}{200mm}                                          
  \caption{HATNet rotational variables in the Praesepe. -- Continued}                                           
  \scalebox{0.90}{                                                
  \begin{tabular}{cclccccccccr}                                   
  \hline                                                          
HAT ID/type & 2MASS ID & $\nu$      &  1/T       &   B   &   V       &   r      &    K       & J$-$H      & J$-$K &   A       &  SNR\\
     &          & [d$^{-1}$] & [d$^{-1}$] & [mag] & [mag]     & [mag]    & [mag]      &  [mag]     & [mag] & [mag]     & \\
\hline
HAT-269-0003629 b  & 08402624+191309 &  0.092922 &  0.0011 & 13.899 & 12.875 & 12.481 & 10.461 &  0.505 &  0.627 &  0.0102 & 12.0\\         
HAT-317-0004241 d  & 08404439+183923 &  0.093539 &  0.0011 & 13.917 & 12.933 & 12.537 & 10.500 &  0.476 &  0.626 &  0.0187 & 14.5\\         
HAT-317-0004286 b  & 08442031+180259 &  0.091827 &  0.0011 & 13.923 & 12.960 & 12.578 & 10.490 &  0.495 &  0.638 &  0.0070 &  8.4\\         
HAT-269-0003847 a  & 08412258+185602 &  0.093008 &  0.0011 & 13.970 & 12.957 & 12.554 & 10.536 &  0.484 &  0.626 &  0.0273 & 19.9\\         
HAT-269-0003924 d  & 08413902+191556 &  0.094158 &  0.0011 & 14.177 & 13.089 & 12.670 & 10.574 &  0.531 &  0.631 &  0.0092 & 11.2\\         
HAT-269-0003973 b  & 08355988+193132 &  0.077594 &  0.0045 & 14.368 & 13.229 & 12.777 & 10.569 &  0.539 &  0.641 &  0.0079 &  9.1\\         
HAT-269-0004092 a  & 08415884+200627 &  0.087139 &  0.0011 & 14.279 & 13.160 & 12.746 & 10.599 &  0.526 &  0.644 &  0.0141 & 11.1\\         
HAT-318-0003767 d  & 08463355+181409 &  0.086355 &  0.0011 & 14.282 & 13.205 & 12.770 & 10.597 &  0.528 &  0.653 &  0.0195 & 13.3\\         
HAT-269-0004164 d  & 08401345+194643 &  0.084119 &  0.0011 & 14.318 & 13.200 & 12.758 & 10.640 &  0.543 &  0.642 &  0.0138 & 13.3\\         
HAT-269-0004180 d  & 08323341+200448 &  0.082029 &  0.0045 & 14.330 & 13.208 & 12.787 & 10.627 &  0.548 &  0.650 &  0.0164 & 11.9\\         
HAT-269-0004361 b  & 08340356+194742 &  0.090750 &  0.0045 & 14.511 & 13.362 & 12.893 & 10.678 &  0.505 &  0.647 &  0.0098 & 10.0\\         
HAT-269-0004323 d  & 08435672+194332 &  0.086016 &  0.0011 & 14.376 & 13.243 & 12.818 & 10.527 &  0.547 &  0.709 &  0.0169 & 12.7\\         
HAT-270-0003419 a  & 08463304+185424 &  0.070295 &  0.0011 & 14.300 & 13.232 & 12.813 & 10.619 &  0.529 &  0.682 &  0.0173 & 17.8\\         
HAT-269-0004069 a  & 08350805+195925 &  0.079749 &  0.0045 & 14.565 & 13.332 & 12.815 & 10.447 &  0.640 &  0.720 &  0.0095 &  9.6\\         
HAT-269-0004532 b  & 08431522+200356 &  0.091513 &  0.0011 & 14.614 & 13.447 & 12.983 & 10.683 &  0.560 &  0.676 &  0.0163 & 16.0\\         
HAT-317-0005295 a  & 08405865+184030 &  0.085103 &  0.0011 & 14.936 & 13.723 & 13.196 & 10.488 &  0.568 &  0.765 &  0.0152 & 12.1\\         
HAT-269-0004773 a  & 08414818+192731 &  0.086184 &  0.0011 & 14.852 & 13.644 & 13.151 & 10.732 &  0.532 &  0.679 &  0.0170 & 15.0\\         
HAT-269-0004680 b  & 08434356+190433 &  0.093232 &  0.0011 & 14.488 & 13.416 & 12.976 & 10.787 &  0.552 &  0.649 &  0.0062 &  8.0\\         
HAT-269-0004793 d  & 08422008+190905 &  0.083953 &  0.0011 & 14.673 & 13.486 & 13.014 & 10.696 &  0.546 &  0.698 &  0.0072 &  8.0\\         
HAT-317-0005320 d  & 08400984+180550 &  0.082000 &  0.0011 & 14.578 & 13.457 & 12.983 & 10.679 &  0.580 &  0.693 &  0.0101 & 11.0\\         
HAT-269-0004728 d  & 08373624+191554 &  0.083522 &  0.0011 & 14.626 & 13.481 & 13.005 & 10.757 &  0.566 &  0.669 &  0.0157 & 11.5\\         
HAT-269-0004769 a  & 08383723+190116 &  0.077810 &  0.0011 & 15.001 & 13.775 & 13.244 & 10.613 &  0.592 &  0.736 &  0.0107 & 10.7\\         
HAT-269-0004494 b  & 08364957+193323 &  0.081739 &  0.0046 & 15.403 & 14.013 & 13.370 & 10.285 &  0.686 &  0.851 &  0.0191 & 13.5\\         
HAT-269-0004958 a  & 08411992+193804 &  0.081357 &  0.0011 & 14.751 & 13.555 & 13.080 & 10.762 &  0.559 &  0.693 &  0.0109 & 10.4\\         
HAT-269-0004816 b  & 08404426+202818 &  0.078254$\times3$ &  0.0192 & 15.672 & 14.254 & 13.650 & 10.402 &  0.663 &  0.838 &  0.0692 &  6.1\\
HAT-269-0004795 b  & 08411319+193234 &  0.078417$\times4$ &  0.0011 & 15.670 & 14.189 & 13.565 & 10.347 &  0.703 &  0.864 &  0.0370 & 16.8\\
HAT-317-0005953 d  & 08344714+180116 &  0.086052 &  0.0045 & 14.829 & 13.678 & 13.188 & 10.830 &  0.571 &  0.697 &  0.0165 & 11.8\\         
HAT-269-0005580 b  & 08372638+192912 &  0.081849 &  0.0011 & 15.111 & 13.831 & 13.303 & 10.826 &  0.531 &  0.730 &  0.0135 & 10.8\\         
HAT-269-0005408 d  & 08390411+193121 &  0.080736$\times2$ &  0.0011 & 14.995 & 13.772 & 13.267 & 10.857 &  0.575 &  0.705 &  0.0088 &  5.9\\
HAT-269-0005362 a  & 08385722+201053 &  0.083890 &  0.0011 & 15.184 & 13.924 & 13.361 & 10.716 &  0.619 &  0.764 &  0.0204 & 13.6\\         
HAT-269-0005801 a  & 08385833+193649 &  0.073795 &  0.0011 & 14.899 & 13.771 & 13.296 & 11.037 &  0.537 &  0.665 &  0.0125 & 11.9\\         
HAT-269-0005589 d  & 08331762+192550 &  0.079399 &  0.0045 & 15.078 & 13.827 & 13.305 & 10.859 &  0.597 &  0.728 &  0.0178 & 10.2\\         
HAT-317-0006380 d  & 08433105+183254 &  0.077285 &  0.0011 & 15.209 & 13.980 & 13.431 & 10.741 &  0.617 &  0.783 &  0.0076 &  8.1\\         
HAT-269-0006020 d  & 08423700+200831 &  0.081840 &  0.0011 & 15.201 & 13.961 & 13.410 & 10.871 &  0.598 &  0.763 &  0.0261 & 14.1\\         
HAT-269-0006002 d  & 08382963+195145 &  0.081521 &  0.0011 & 15.308 & 14.018 & 13.469 & 10.931 &  0.607 &  0.739 &  0.0200 & 15.7\\         
HAT-317-0006654 d  & 08365374+182945 &  0.076722 &  0.0046 & 15.218 & 13.985 & 13.437 & 10.947 &  0.622 &  0.725 &  0.0172 &  9.4\\         
HAT-269-0006168 d  & 08442652+194735 &  0.071629 &  0.0011 & 15.466 & 14.125 & 13.574 & 10.908 &  0.628 &  0.767 &  0.0182 &  9.9\\         
HAT-317-0006713 d  & 08443613+183557 &  0.079219 &  0.0011 & 15.905 & 14.518 & 13.921 & 10.655 &  0.699 &  0.860 &  0.0236 &  9.8\\         
HAT-269-0006357 b  & 08414934+191147 &  0.070891 &  0.0011 & 15.864 & 14.466 & 13.846 & 10.834 &  0.636 &  0.815 &  0.0154 &  8.3\\         
HAT-269-0006450 b  & 08421664+200532 &  0.070833 &  0.0011 & 15.498 & 14.187 & 13.605 & 10.989 &  0.607 &  0.757 &  0.0183 &  9.4\\         
HAT-269-0006795 b  & 08441324+184911 &  0.077282 &  0.0011 & 15.431 & 14.181 & 13.612 & 10.941 &  0.616 &  0.806 &  0.0213 & 15.1\\         
HAT-269-0006689 d  & 08424596+211616 &  0.070576$\times4$ &  0.0192 & 15.650 & 14.326 & 13.765 & 10.919 &  0.637 &  0.811 &  0.0311 &  5.2\\
HAT-269-0006856 b  & 08400070+191834 &  0.070008 &  0.0011 & 15.805 & 14.427 & 13.829 & 10.918 &  0.648 &  0.826 &  0.0212 & 11.2\\         
HAT-269-0007142 b  & 08313281+210128 &  0.052230$\times2$ &  0.0192 & 15.881 & 14.573 & 13.979 & 11.118 &  0.636 &  0.767 &  0.0658 &  7.8\\
HAT-269-0007074 b  & 08391580+200414 &  0.065576$\times2$ &  0.0048 & 16.432 & 14.946 & 14.292 & 10.867 &  0.686 &  0.871 &  0.0369 &  8.8\\
HAT-269-0007168 a  & 08411541+200216 &  0.073167 &  0.0011 & 15.563 & 14.230 & 13.669 & 11.022 &  0.657 &  0.811 &  0.0248 & 12.2\\         
HAT-269-0007109 b  & 08365680+190528 &  0.067361 &  0.0047 & 16.115 & 14.659 & 14.047 & 10.974 &  0.680 &  0.829 &  0.0222 &  9.7\\         
HAT-269-0006935 b  & 08312987+202437 &  0.089114$\times4$ &  0.0208 & 15.766 & 14.351 & 13.729 & 10.767 &  0.725 &  0.907 &  0.0459 &  5.3\\
HAT-317-0008551 d  & 08405531+183459 &  0.080526 &  0.0011 & 15.838 & 14.507 & 13.911 & 11.128 &  0.660 &  0.796 &  0.0165 & 10.8\\         
HAT-270-0006183 b  & 08534667+191814 &  0.072508 &  0.0065 & 16.282 & 14.798 & 14.207 & 11.165 &  0.660 &  0.808 &  0.0278 &  6.3\\         
HAT-269-0008437 b  & 08363642+191106 &  0.074895 &  0.0047 & 16.180 & 14.719 & 14.107 & 11.199 &  0.600 &  0.820 &  0.0254 &  9.6\\         
HAT-269-0008217 a  & 08314045+194754 &  0.057273$\times2$ &  0.0048 & 16.580 & 15.098 & 14.495 & 11.027 &  0.694 &  0.890 &  0.0615 &  9.4\\
HAT-269-0008376 d  & 08305102+192108 &  0.070732 &  0.0047 & 16.101 & 14.752 & 14.176 & 11.193 &  0.661 &  0.829 &  0.0194 &  9.2\\         
HAT-317-0010700 a  & 08415228+180306 &  0.064425 &  0.0012 & 16.451 & 15.091 & 14.470 & 11.337 &  0.661 &  0.833 &  0.0236 & 10.3\\         
HAT-269-0009602 a  & 08394103+195928 &  0.059496$\times4$ &  0.0061 & 17.053 & 15.339 & 14.736 & 11.323 &  0.679 &  0.854 &  0.0611 &  7.2\\
HAT-318-0009219 a  & 08492676+183119 &  0.055142$\times2$ &  0.0012 & 16.819 & 15.396 & 14.771 & 11.394 &  0.656 &  0.847 &  0.0742 & 11.2\\
HAT-317-0011466 a  & 08412417+181402 &  0.119670$\times4$ &  0.0067 & 17.082 & 15.787 & 15.193 & 11.455 &  0.660 &  0.824 &  0.1061 &  7.1\\
HAT-269-0010318 b  & 08393645+192907 &  0.063811 &  0.0011 & 16.544 & 15.106 & 14.501 & 11.391 &  0.652 &  0.863 &  0.0239 &  8.9\\         
HAT-318-0010036 a  & 08483271+165623 &  0.055839$\times3$ &  0.0065 & 17.240 & 15.759 & 15.200 & 11.468 &  0.625 &  0.860 &  0.0726 &  8.0\\
HAT-317-0012279 b  & 08363256+162302 &  0.063021 &  0.0061 & 16.839 & 15.369 & 14.770 & 11.508 &  0.649 &  0.839 &  0.0388 &  6.3\\         
\hline                                                            
\end{tabular}}                                                    
\end{minipage}                                                    
\begin{flushleft}                                                 
\underline{Notes:}
Variable types based on the classification of their frequency spectra (see 
Sect.~3.3) are shown in the second column. Frequencies flagged by $\times n$ 
indicate that the value shown is $1/n$-th of the most dominant frequency of 
the DFT spectrum. As a proxy of the error of the frequencies, we list the 
inverse of the total time span (T). The B, V, Sloan {\em r}, K, J$-$H and 
J$-$K magnitudes are from the APASS and 2MASS databases, respectively. 
Amplitudes (A) are peak-to-peak values from the 4$^{\rm th}$ order Fourier 
fit to the data without TFA filtering and using the frequencies as given in 
this table. The signal-to-noise ratio (SNR) of the peak frequency is computed 
from the data without TFA filtering and refers to the $[0,0.5]$~d$^{-1}$ band.                                               
\end{flushleft}                                                   
\end{table*}                                                      
%

After multiple screenings of all the stars analyzed, we end up with
$180$ rotational variables (see Table~\ref{data-rot-var}). The
positions of these stars on the r[mag]--period plane are shown in
Fig.~\ref{rmag-period-2rvp}.  We see that indeed, the relation is
fairly tight, although there are some $13$ outliers (arbitrarily
defined as those with relative distances from the ridge greater than
$20$~\% -- i.e., $|\nu/\nu_{\rm ridge}-1| > 0.2$).  Except perhaps for
a few low-SNR variables, all are very secure detections.  None of them
have other frequencies in the $[0,50]$~d$^{-1}$ band (except for
possible harmonic components). Some of them at the fainter (redder)
end may well belong to the group of stars that have not yet converged
to a tight rotational sequence. An additional source of ambiguity in 
this diagram is the reliability of the associated magnitude/color 
values. These might be strongly affected by blending in the more crowded 
region of the cluster. We also mention that except for two 
(HAT-269-0000761 and HAT-269-0000794) of the $24$ spectroscopic binaries 
common with the list of \cite{mermilliod2009}, all follow the overall 
r[mag]--period relation shown in Fig.~\ref{rmag-period-2rvp}. 

It is important to note that we also detected the rotation of the two 
planet hosts. The ratios of the orbital to the rotational periods are 
non-commensurate in both systems. For Pr0201=HAT-269-0000805 
$P_{\rm orb}/P_{\rm rot}=0.787$ whereas for Pr0211=HAT-269-0002316 
this ratio is $0.270$. This further supports the planetary nature of 
these systems, since the periods of the radial velocity signals are not 
integer multiples/fractions of the rotational periods as one might expect 
if the radial velocity variations were caused by stellar activity. 

In comparison with similar plots from other clusters, it is seen that we 
miss the faint end of the progression, where the lack of magnetic braking 
shows up as a suddenly increased scatter in the periods. Our survey (similarly 
to that of \citealt{delorme2011}) is too shallow to cover this rather faint 
magnitude range. However, the PTF project covered just this range for this 
cluster, and, indeed, found $40$ variables with a wide period range 
of $0.5$ to $36$~days \citep{agueros2011}. Similarly, the deep survey by 
\cite{scholz2011} resulted in the discovery of $26$ fast rotating 
($P_{\rm rot} < 2.5$~d) objects in the very low ($M < 0.3 M_{\odot}$) mass 
regime. The recent analysis by \cite{mcquillan2013a} confirms this wide 
period distribution also for the Kepler field M stars. Toward the blue end, 
on the other hand, due to the field overlaps, our survey is sensitive 
to periodic variations down to a few mmag. This leads to a good coverage of 
the progression in this regime, where the roughly linear color dependence 
becomes nonlinear and the rotational periods get rather short. For other 
clusters, a more detailed exploration of this regime is under way by the 
analysis of the data from the Kepler satellite (see \citealt{meibom2011} 
on NGC~6811 and \citealt{balona2013} on NGC~6866). 

It is interesting to compare the list of the $52$ rotational variables
published by \cite{delorme2011} from the SuperWASP survey, covering
basically the same color and magnitude range as HATNet. We find
matching HATNet targets for all $52$ stars, $7$ of which are not
included in the \cite{kraus2007} list of cluster members. Our data do
not yield significant signals for $13$ stars. (Although some of them
may be present in the original EPD data without TFA filtering, we do
not consider them as valid detections until they are confirmed by the
analysis made on the TFA-filtered data.) For the remaining $39$ mutual
detections we have good agreement between the SuperWASP and HATNet
frequencies (with relative frequency differences of 
$|\nu_{\rm HAT}/\nu_{\rm WASP} - 1|$ within $10$\% for the mutually detected
$39$ stars and within $3$\% for $75$\% of these stars).

The two highlighted planet hosts (see the bigger filled circles) basically 
follow the overall rotational trend, but the fainter one seems to fall a bit 
on the short-period side. In Sect.~4 we further discuss the relation of 
these and other planet hosts to the overall rotational patterns in different 
clusters. 

For an easier reference to the ridge values in the various colors, we fit 
$m$-th order polynomials to the respective color-period relations
 
\begin{equation}
P = \sum_{i=1}^{m+1} a_{i-1}x^{i-1} \hskip 2mm ,
\end{equation}

\noindent
where $x=k({\rm c}-<c>)$, $k=2/(\rm{cmax}-\rm{cmin})$, 
$<c>=(\rm{cmax}+\rm{cmin})/2$ and 
$\rm{cmin}$, $\rm{cmax}$, respectively, refer to the (min,max) values of 
the given dereddened color `$c$' in the region we fit the periods. 
Dereddening is performed with $E(B-V)=0.027$ and standard relative 
reddening coefficients (e.g., \citealt{yuan2013}). For V, B$-$V and 
Sloan {\em r} the order of the polynomial ($m$) is equal to $4$, whereas 
for the 2MASS colors we choose $m=3$, because of the lower stability of 
the fit for these colors at higher polynomial orders. In finding the best 
fit to the ridge, we employ sigma clipping at the level of $2.5\sigma$. 
The resulting coefficients are listed in Table~\ref{polfit-rot-var}. 

%
\begin{table}
 \begin{minipage}{140mm}
  \caption{Polynomial fits to the ridge values.}
  \label{polfit-rot-var}
  \scalebox{0.80}{
  \begin{tabular}{rrrrrrrr}
  \hline
   color  & $a_0$  & $a_1$  & $a_2$ & $a_3$ & $a_4$ & cmin & cmax \\
  \hline
   V  & $10.724$ & $4.356$ & $ 3.011$ & $3.240$ & $-3.942$ & $9.941$ & $15.787$ \\
 B-V  & $11.528$ & $4.934$ & $ 1.277$ & $1.446$ & $-2.897$ & $0.483$ & $ 1.714$ \\ 
   r  & $10.006$ & $3.132$ & $ 1.147$ & $3.483$ & $-2.233$ & $9.906$ & $14.218$ \\
   K  & $ 9.579$ & $4.633$ & $-0.755$ & $4.083$ & $ 0.000$ & $8.580$ & $11.505$ \\
 J-H  & $10.380$ & $3.917$ & $-0.240$ & $2.096$ & $ 0.000$ & $0.180$ & $ 0.722$ \\
 J-K  & $10.447$ & $2.975$ & $ 0.675$ & $4.020$ & $ 0.000$ & $0.247$ & $ 0.903$ \\
\hline
\end{tabular}}
\end{minipage}
\begin{flushleft}
\underline{Notes:} 
See Eq.~(2) and the subsequent explanation for the use of the data in this 
table. By omitting outliers at the $3\sigma$ level, the above regressions 
fit the $180$ rotational variables with standard deviations of $0.7$, $0.9$, $0.7$, 
$1.0$, $1.2$ and $1.2$~days for V, B$-$V, {\em r}, K, J-H and J-K, respectively. 
The number of the corresponding outliers are $11$, $7$, $11$, $8$, $5$ and $3$. 
\end{flushleft}
\end{table}

%
\subsection{Compatibility with the observed rotational velocities}
\begin{figure}
 \vspace{0pt}
 \includegraphics[angle=-90,width=85mm]{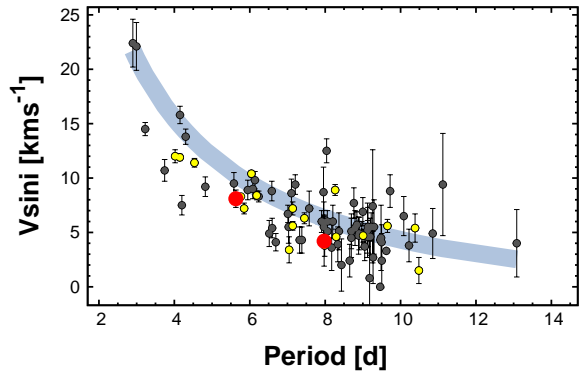}
 \caption{Photometric periods versus observed rotational velocities by 
         \protect\cite{mermilliod2009} for the objects in common with our 
	 list of rotational variables (see Table~\ref{data-rot-var}). 
	 The RV planet hosts are shown by the bigger filled circles, 
	 the spectroscopic binaries from \protect\cite{mermilliod2009} by 
	 the ones with lighter shade. The thick continuous line is 
	 drawn by using Eq.~(3). The line covers the stellar rotational 
	 velocities derived from the observed rotational periods and 
	 stellar radii given by model isochrones.}
\label{vrot-period}
\end{figure}

If spectroscopic $V_{\rm rot}\sin i$ values are available, it is
customary to compare these values with the derived rotational
periods. We use the data from the spectroscopic survey of open
clusters by \cite{mermilliod2009}.  By cross-correlating the data
published for Praesepe with our rotational variable list, we find $88$
matches. These stars are plotted in Fig.~\ref{vrot-period}. For the
computation of the expected rotational rates we estimate the radii of
the stars by using the Padova isochrones\footnote{see 
\cite{bressan2012} and {\bf http://stev.oapd.inaf.it/cmd}} for the 
age of $590$~Myr with solar-scaled composition at $Z=0.02$. For the 
overall cluster reddening we set $E(B-V)=0.027$\,mag and for the 
distance modulus we use $6.30$\,mag. These values are taken from
\cite{fossati2008}, \cite{boesgaard2013}, \cite{taylor2006} and
\cite{vanleeuwen2009} and are in the overall ranges of the most
frequently quoted parameters for this cluster.

Although the isochrone corresponding to the above parameters fits the 
($J-K$, $K$) diagram quite well in the $J-K$ regime between $0.2$ and 
$0.8$, it fails to reproduce the break at $J-K\sim0.85$. As discussed by 
\cite{khalaj2013}, other isochrones fit this particular part better 
but are less successful in reproducing the hotter part of the main 
sequence. Since we are focusing on stars between A and M spectral types, 
and our experiments with other isochrones led to similar results to 
those of the Padova isochrones, we decided to use these latter ones in 
the estimation of the stellar radii. We found that the computed stellar 
rotational velocities can be well fitted by the following simple 
formula
  
\begin{equation}
V_{\rm rot} = 57.0P_{\rm rot}^{-(1.0+0.03(P_{\rm rot}-6.0))} \hskip 2mm ,
\end{equation}

\noindent
where $P_{\rm rot}$ is the measured rotational period (in [days]). The 
thickness of the line shown in Fig.~\ref{vrot-period} covers the range of 
scatter due to observational errors in the 2MASS colors. It is important 
to note that the above formula can be employed only on this cluster, since 
it depends on the age and chemical composition, specific to this cluster. 

We think that the agreement between the observed and expected rotational
velocities is reasonably good. For ideal (errorless) observed rotational 
velocities we should see the `theoretical' equatorial velocities (thick gray 
line in Fig.~\ref{vrot-period}) as the upper envelope of the observed values. 
We see that some stars are above this line. Several of them have reasonably 
large error bars to consider their outlier status insignificant, but some of 
them may pose some concern, such as HAT-269-0001124 at the period of $8.04$~d 
with small error and well-established period.

%
%

%
\subsection{Color -- amplitude dependence} 
It is important to investigate if the amplitude of the observed 
variability has any tendency of a correlated variation with some other 
properties of the star. Such a correlation might be expected from  
the underlying dependence of stellar activity on the same parameters 
(e.g., \citealt{mamajek2008}). Because the observed photometric amplitude 
is a multivariable function of the complicated physical processes 
(rotation, local and global magnetic field variations, convection, 
turbulence, etc.) involved in the spot generation and also of the 
specific geometric configuration, we expect rather fuzzy relations between 
the amplitudes and some other observable parameters. Indeed, results 
published so far show that the relation is fairly vague, especially if 
we are aimed at a more comprehensive analysis (see \citealt{saar2011} 
for a step in this direction).    

\begin{figure}
 \vspace{0pt}
 \includegraphics[angle=-90,width=85mm]{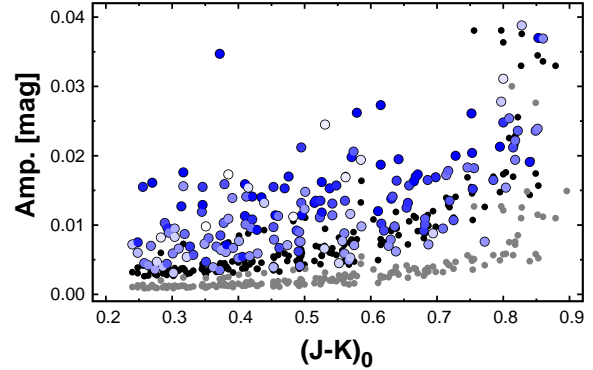}
 \caption{Variation of the peak-to-peak amplitudes (see 
          Table~\ref{data-rot-var}) as a function of the 2MASS 
	  $(J-K)_0$ color (large shaded circles, with shading 
	  code from $SNR\le4$ [white] to $SNR\ge14$ [deep blue/gray]). 
	  For reference, the $5$ and $15\sigma$ limits of signal detection 
	  (assuming white noise) are also shown on a star-by-star basis 
	  by gray and black filled circles, respectively. In order to 
	  scale the figure to the bulk of the sample, we left out a 
	  handful of stars with amplitudes larger than $0.04$~mag. The 
	  apparent trend in the amplitudes is most likely associated with 
	  the higher detection limit toward redder colors (i.e., fainter 
	  stars).}
\label{color-amp}
\end{figure}

To add further piece of information to this issue, by using the data 
of Table~\ref{data-rot-var}, we show the variation of the peak-to-peak 
amplitudes as a function of $(J-K)_0$ in Fig.~\ref{color-amp}. The 
dependence is similar to most of the published results; redder (fainter, 
longer-period) stars tend to have larger amplitudes. We note that employing 
the Rossby number (the ratio of the rotation period to the convective 
turnover time -- see \citealt{noyes1984}), advocated by some studies (e.g., 
\citealt{hartman2009}, \citealt{walkowicz2013}), yields also consonant 
result with similar scatter. Because of the data are more noisy and may 
contain less data points at the red (faint) end, we must address the question 
if the observed trend is the result of noise-limited sampling. By using the 
simple assumption of Gaussian white noise, the Fourier amplitudes follow 
a frequency-independent $\chi$ distributions, therefore we can compute the 
average expected Fourier amplitude as $<A_{\rm noise}>=\sqrt{\pi/N}\sigma$. 
Here $N$ is the number of data points and $\sigma$ is the (assumed to be) 
uniform point-by-point standard deviation of the data. We estimate the 
expected signal amplitudes at the detection level of $k\sigma$ as 
$k<A_{\rm noise}>$. It is seen that the higher significance lower limits 
(black dots) nicely follow the lower boundary of the amplitudes of the 
detected variables. The lower significance lower limits (gray dots) show 
the same trend but rendering the suspected trend significant. Considering 
that our assumption on the noise properties is a gross simplification of 
the real data setting with gaps, systematics and associated colored noise, 
we incline to conclude that most of the observed trend comes from the 
lower signal detection efficiency at the high-noise end. The similarity 
of the trend of the statistical limit to that of the lower boundary of 
the variables also supports this conjecture. 

%
%

%
\subsection{The four types of variability}
One particular problem with the photometric identification of the rotating 
variables is that the corresponding time series are non-stationary due to 
the finite life time and migration of the spots to different latitudes, 
where differential rotation leads to a frequency drift in the observed signal. 
In spite of this, it is expected that some level of coherence prevails 
and leads to a broad, but still well-defined peak in the frequency spectrum. 

Although the peak profiles can supply valuable information on spot life 
times and migration/differential rotation, the current data are still not 
suitable for this kind of deep analysis, due to the interfering noise 
and gaps in the sampling. Nevertheless, some basic properties are easy to  
recognize and one can classify the observed frequency spectra based on 
these simple characteristics. Although this classification is phenomenological 
in nature, it adds further solid information to the occurrence rates of the 
most common types of rotation-induced variability. 

\begin{figure}
\centering
  \begin{tabular}{@{}cccc@{}}
    \includegraphics[width=.22\textwidth]{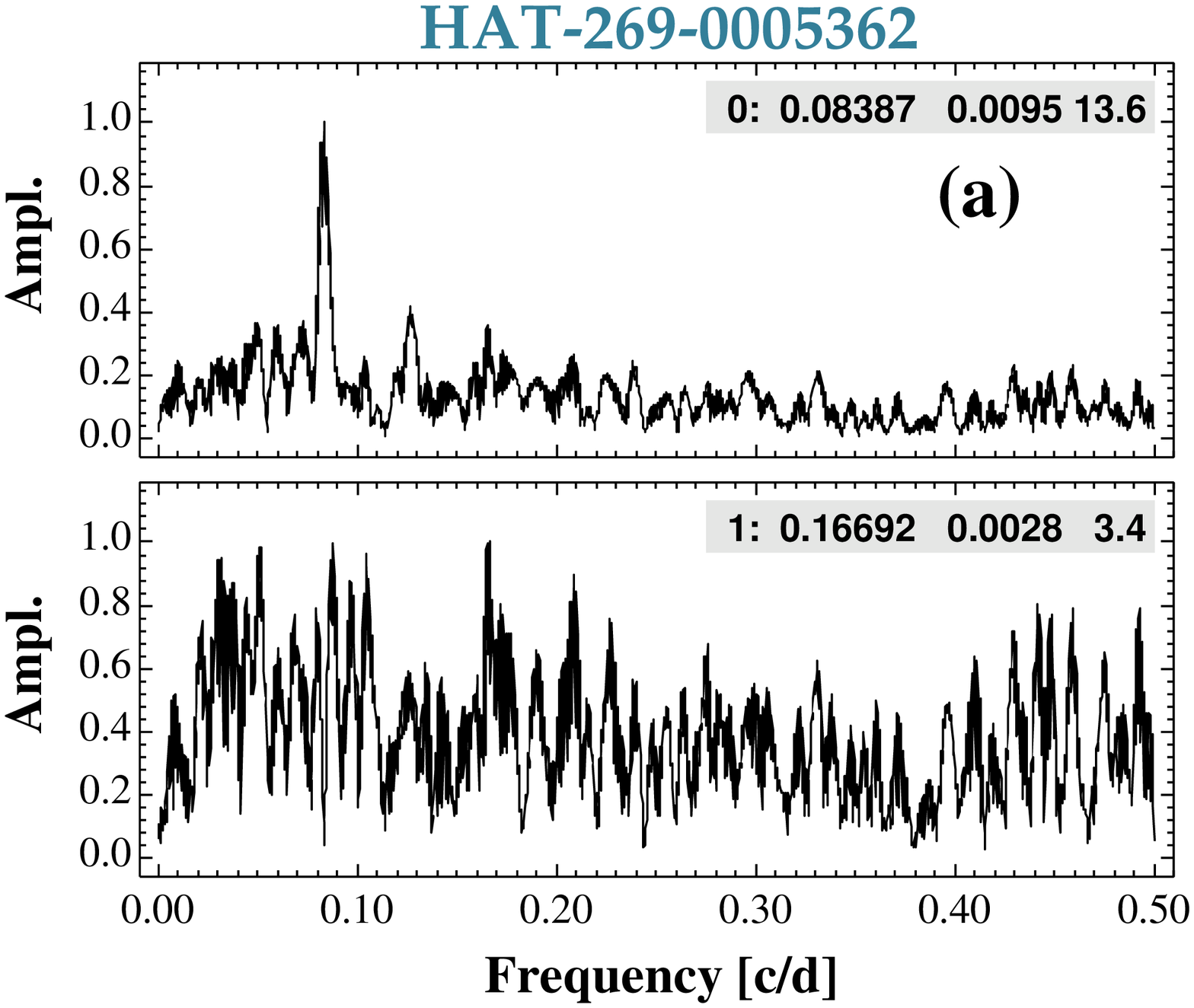} &
    \includegraphics[width=.22\textwidth]{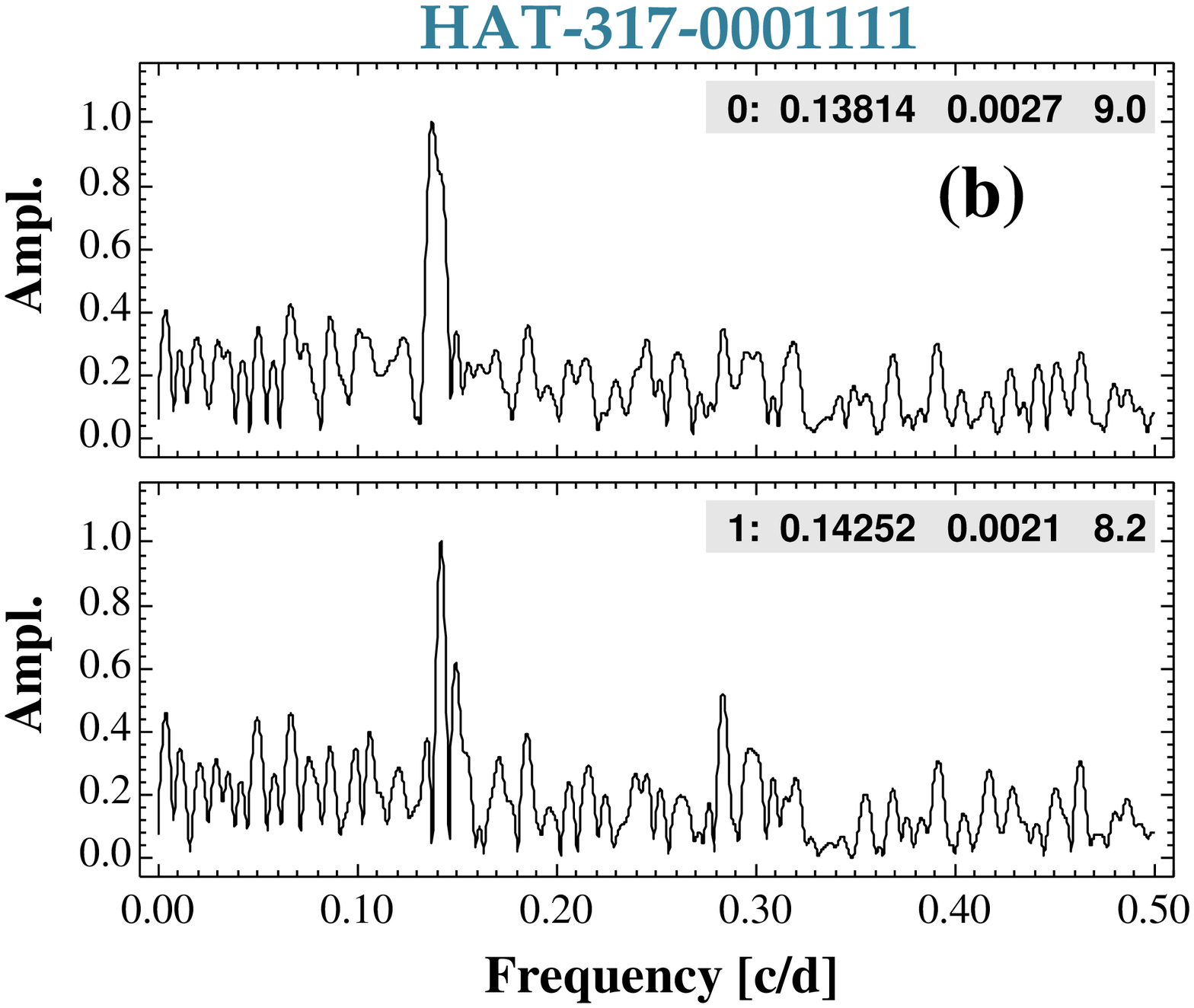} \\
    \includegraphics[width=.22\textwidth]{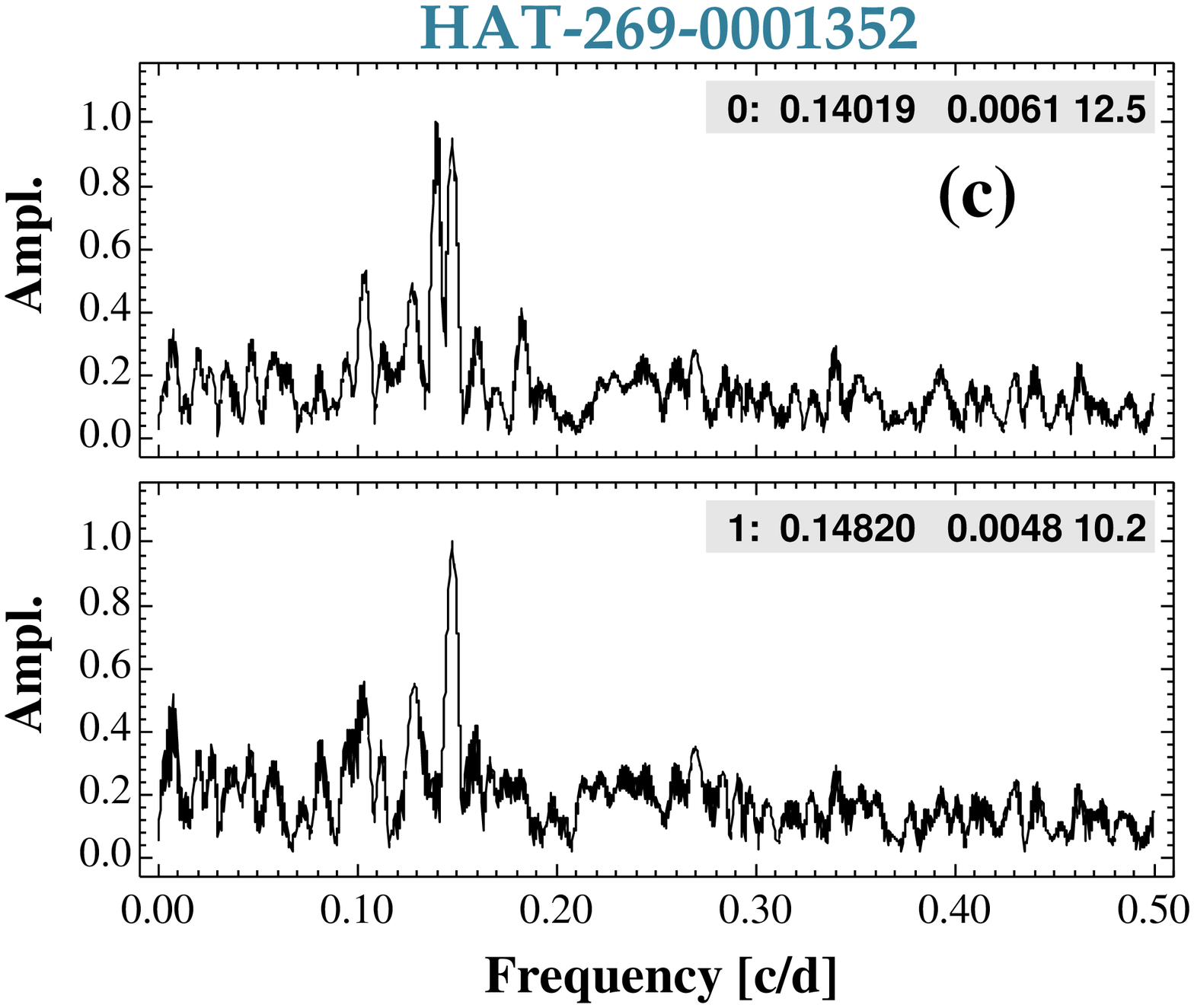} &
    \multicolumn{2}{c}{\includegraphics[width=.22\textwidth]{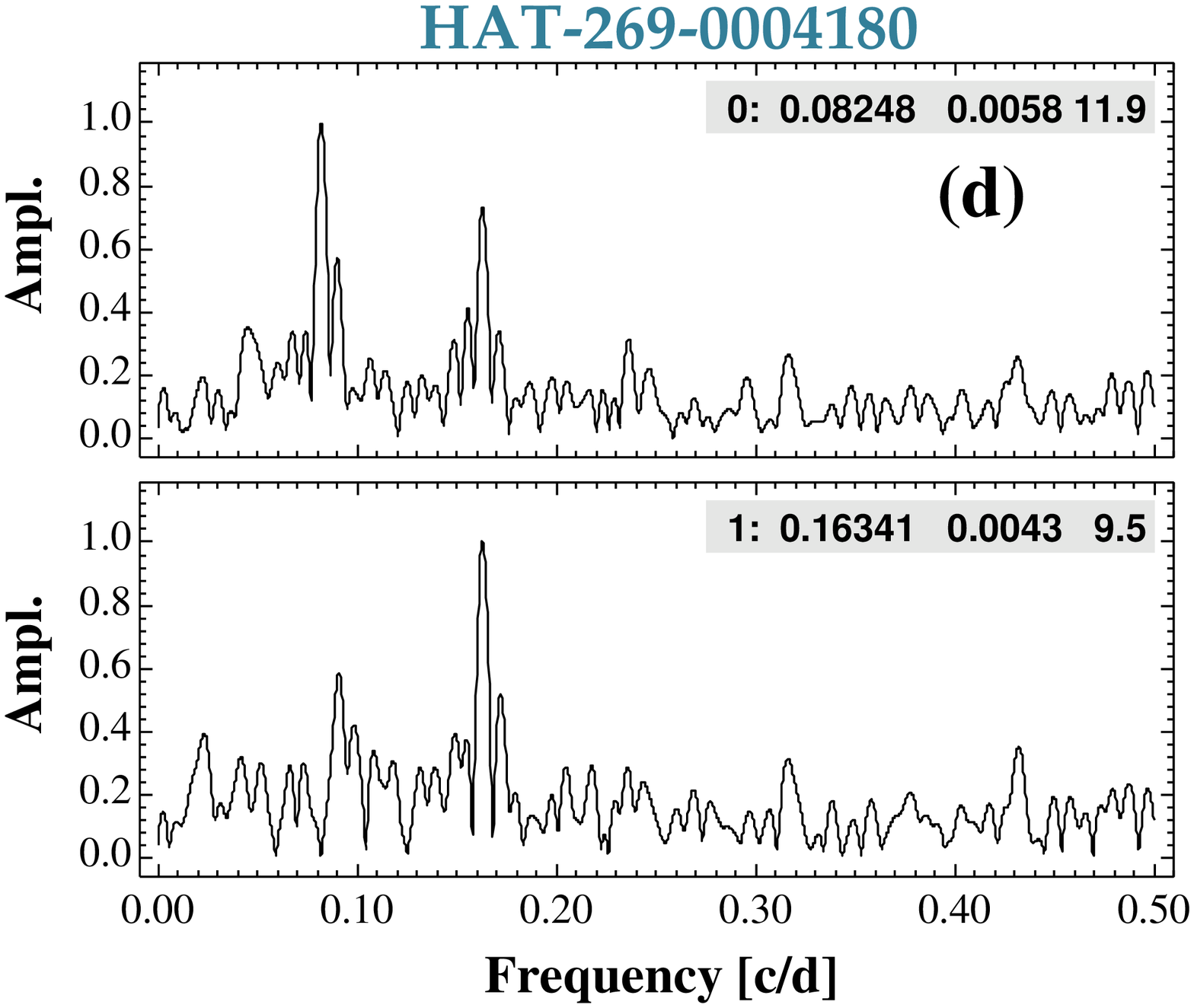}}
  \end{tabular}
  \caption{Examples for the four types of frequency spectra of the rotational 
  variables as discussed in Sect.~3.3. In each sub-figure the upper and lower 
  panels show, respectively, the frequency spectra of the data and the spectra 
  obtained after the first pre-whitening (with the peak frequency shown on the 
  upper panel). The insets display the pre-whitening order, the peak frequency,  
  the amplitude and the SNR of the peak. The spectra are normalized to the 
  highest peak in each panel.}
\label{4types-spectra}
\end{figure}

Upon examining the frequency spectra (with some representative examples 
shown in Fig.~\ref{4types-spectra}), we find that all $180$ stars classified 
as rotational variables fall in one of the following categories.\footnote{In 
addition to the possibility of overlapping properties, as always, there is 
the issue of noise. This may shift the actual statistics, depending on where 
the level of significance is set (which may be an issue here, since the noise 
is usually colored, leading to frequency-dependent detection significance). 
Here we try to avoid any dubious cases and rely on well-established detections.}

%

\begin{enumerate}
\renewcommand{\theenumi}{(\alph{enumi})} 
\item
These variables apparently have monoperiodic sinusoidal light variation 
(i.e., no residual power left near the main peak after pre-whitening). 
We have $23$ such cases in the present sample, which constitutes $13$\% 
of the rotating variables. 
\item
There are $71$ variables (i.e., $39$\% of the full sample) with a single 
but unstable sinusoidal component. Because of the frequency/amplitude 
drifts, these variables have inherently non-discrete Fourier spectra. 
Therefore, they show residual power close to the peak frequency even 
after several pre-whitening cycles (based on the piece by piece signal 
removal, assuming stationarity for each component). 
\item
In $13$ infrequent cases we can observe two peaks separated by a few times 
of the overall line width (i.e., by $\Delta\nu=1/220$d$^{-1}$, where the basic 
time span of $220$~days is determined by the compact/$\sim$continuous  
time base of two of our fields, \#$269$ and $317$). The frequency distances 
are between  $\sim 0.007$d$^{-1}$ and $\sim 0.020$d$^{-1}$ (with the exception 
of HAT-270-0000714, that has frequency components separated by 
$\sim 0.032$d$^{-1}$). The most common explanation of these separate components 
is that they are associated with the different latitudes of the stellar surface, 
that, due to differential rotation, carry the respective spots with different 
periods.  We note that even the largest frequency separation is allowed by 
current models of differential rotation (e.g., \citealt{kitchatinov2012}). 
These $13$ stars constitute $7$\% of the full sample.  
\item
As already indicated in Fig.~\ref{rmag-freq-all}, most of the frequencies 
with sufficient significance that do not sit on the main rotational ridge 
in the color/magnitude -- period plot belong to the first harmonic of the 
rotational frequency. The first harmonic may have larger or smaller amplitude 
than that of the rotational frequency. There are altogether $73$ stars (i.e., 
$41$\% of the sample) that have residual power either at the corresponding 
1st harmonic or at a sub-harmonic frequency. The effect of the harmonics may 
show up either as a non-sinusoidal distortion or as a double hump in the light 
curve, indicating more complicated spot distribution in the latter case. 
\end{enumerate}

\begin{figure}
 \vspace{0pt}
 \includegraphics[angle=-90,width=85mm]{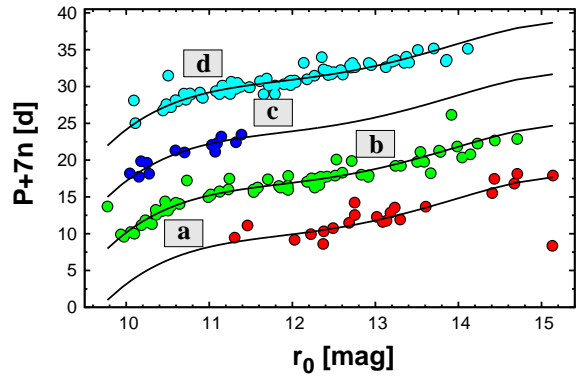}
 \caption{Rotational variables plotted in the dereddened Sloan {\em r} 
          magnitude -- period plane and separated according to the classification 
	  presented in Sect.~3.3. For better visibility, the various classes are 
	  shifted by 7~days in the vertical directions (with type (a) left at the 
	  original periods).}
\label{var-class}
\end{figure}

It maybe useful to investigate how the above type of formal classification 
relates to other parameters, in particular to the magnitudes or colors. In 
Fig.~\ref{var-class} we plot the above classes separately in the $r_0$~--~$P$ plane. 
Interestingly, not all types of variables are distributed evenly in this plane. 
Although the particular distributions are related to the types with relatively low 
number of members, we think that for type (c) (variables with two closely spaced 
frequencies) and maybe also for type (a) (those with constant amplitude sinusoidal 
variations) we may state that they are preferably found at high and low/mid 
luminosities, respectively. Although we cannot offer an explanation for this 
distribution pattern at this moment, it is important to note that for short 
rotation periods (i.e., higher luminosities) differential rotation results in 
a larger difference in the rotational periods between the pole and the equator 
(e.g., \citealt{kitchatinov2012}). Therefore, for hotter (fater rotating) stars, 
frequency splitting due to differential rotation might be more easily observed 
and disentangled from line profile broadening due to the nonstationary nature 
of the spots. On the other hand, the observed frequency spacings do not increase 
with the luminosity as one might expected from the result of \cite{kitchatinov2012}.    

\begin{figure*}
 \vspace{0pt}
 \includegraphics[angle=0,width=150mm]{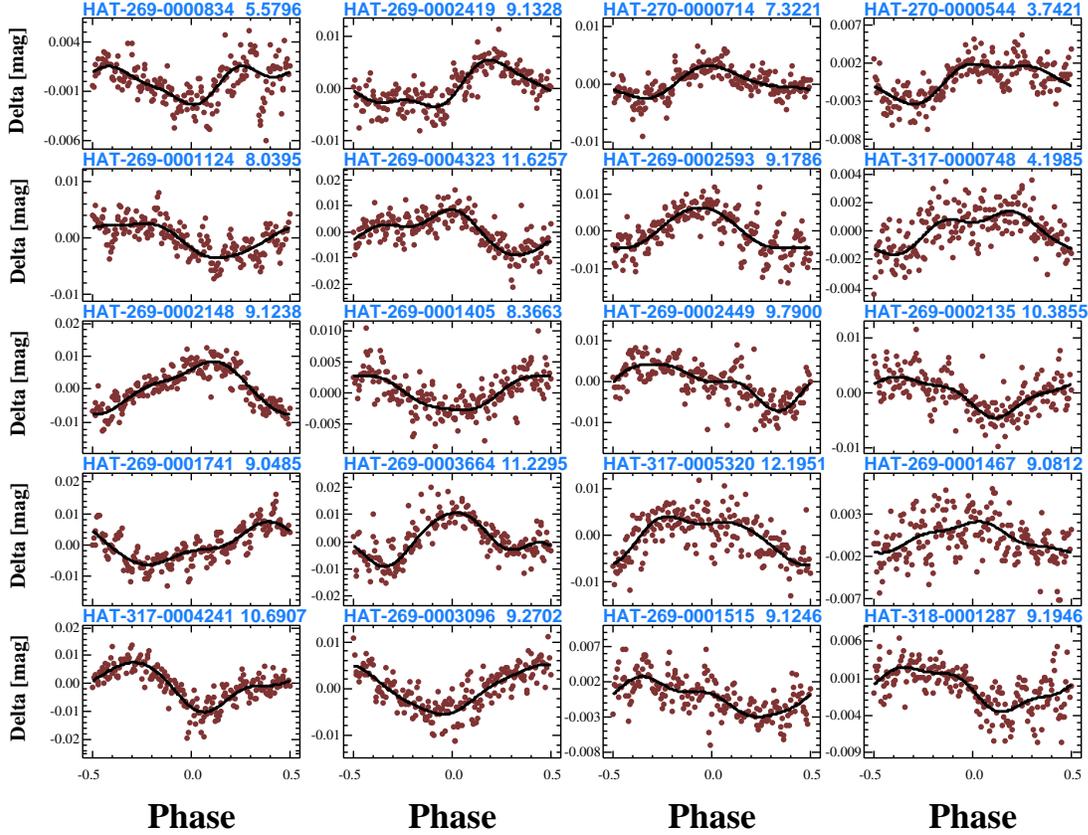}
 \caption{Example of the mid-SNR light curves from the $180$ rotational 
 variables. For better visibility, the light curves are binned in $200$ 
 bins. The corresponding 3rd-order Fourier fits are shown by continuous 
 lines. The folding period is equal to the rotational period 
 (see Table~\ref{data-rot-var}) and shown (in [days]) in the upper right 
 corner of each panel. For this plot we use the original (i.e., EPD, 
 non-TFA-filtered) data. As a result, some of the light curves may contain 
 substantial amount of systematics (e.g., HAT-269-0000834). Phase zero 
 corresponds to HJD~2456000.0 .}
\label{lc-sample}
\end{figure*}

Finally, for a more direct inspection of the rotational variables, we display 
some mid-SNR LCs in Fig.~\ref{lc-sample}. It is seen that non-sinusoidal variations 
are quite common and the LCs are diverse as expected from different geometrical and 
spot configurations.

%
%

%

%
\section{Do close-in planets spin up their host stars?}
For close-in planets it is an important question if tidal interaction has 
any influence on the dynamics of the system, including the rotation of the 
star. For individual systems in the field it is often difficult to study 
this question, since most of the parameters (e.g., age, rotational period) 
necessary to make any reasonably solid statement on this interaction, are 
among the least accurately determined quantities for single free-floating 
stars.   

%
\begin{figure}
 \vspace{0pt}
 \includegraphics[angle=-90,width=85mm]{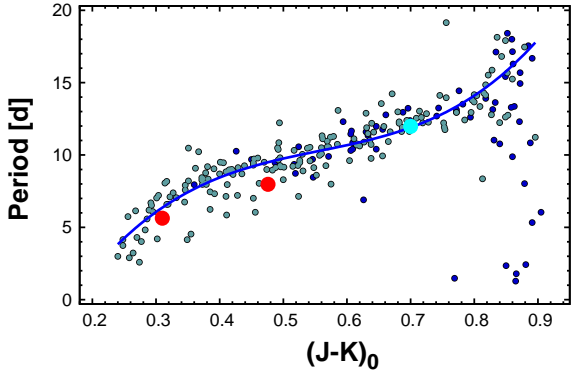}
 \caption{Period--(J$-$K)$_{0}$ plot for Praesepe (light filled 
          circles, this paper) and for the Hyades (dark filled circles, 
	  \protect\citealt{delorme2011}). The RV planets 
	  (\protect\citealt{quinn2012, quinn2014}) are shown by the 
	  larger filled circles; darker shade for Praesepe and lighter 
	  one for Hyades. For reference, the 3rd order polynomial fit 
	  to the Praesepe data is also shown.}
\label{j-k-3rv-planets}
\end{figure}

With the discovery of the first HJs in Praesepe and Hyades
(\citealt{quinn2012, quinn2014}) and quite recently also in M67
\citep{brucalassi2014}, and the first transiting planets in NGC~6811
\citep{meibom2011} it has become possible for the first time to use
well-established cluster ages and relative rotational rates to study if
planets have any systematic effects on the rotation of their host
stars.  Although the sample is still fairly small ($8$ planets, discarding 
the ones with orbital periods longer than $20$ days), 
we may attempt to examine the relative positions of the planet hosts 
of these systems on the corresponding color--period diagrams, and see 
if they show any systematic differences relative to the single (supposedly 
no giant planet host) stars. Unfortunately, M67 is relatively old, and 
there are not enough published rotational periods available (e.g., 
\citealt{martins2011, stassun2002}). The two TEPs in NGC~6811 have longer 
periods, and most probably lower masses (in the range of Neptune), 
therefore, they are probably less relevant in the context of star-planet 
interaction. Fortunately, the remaining three planets in Praesepe and 
Hyades have photometrically determined rotational rates, as do many of 
their cluster-mates.
 
We show the three RV planets together with other known rotational variables 
on the (J$-$K) -- period plane in Fig.~\ref{j-k-3rv-planets}. The rotational 
rates in the two clusters nicely follow the same pattern in the overlapping 
region of the main rotational ridge. This further confirms that the two 
clusters have the same age \citep{delorme2011}. All three planet hosts are 
fairly close to the main ridge but the two in Praesepe may be slightly 
below it. When comparing with a similar plot for the Sloan {\em r} color 
(Fig.~\ref{rmag-period-2rvp}) we see that although the overall topology is 
the same, there are differences, attributed in part to the various noise 
properties of Sloan magnitudes and 2MASS color indices. 

We tested various color combinations and appropriate transformations 
to stellar mass and temperature. They all gave very similar topology, albeit 
with different scatters. The status of both the cooler 
(Pr0211=HAT-269-0002316, $P_{\rm orb}=2.15$~d, $P_{\rm rot}=7.97$~d) and the 
hotter (Pr0201=HAT-269-0000805, $P_{\rm orb}=4.43$~d, $P_{\rm rot}=5.63$~d) 
planet hosts remained basically the same: the hotter one was always closer 
to (and likely to be on) the ridge, whereas the cooler one retained its 
relatively distinct position from the ridge. The planet host in the Hyades 
was also always close to the ridge but usually slightly on the short period 
side. When checking a similar plot of \cite{meibom2011} for NGC~6811 (their 
Fig.~1b), we see that the two planet hosts have also slightly shorter periods 
than most of the stars at the same colors.    

As far as Pr0211, the cooler planet host in Praesepe, is concerned, we 
find that the distance from the cubic polynomial fit to the ridge (see 
Sect.~3 and Table~\ref{polfit-rot-var}) varies between $-1.96$ and 
$-1.62$~days, depending on the color combination used.\footnote{For 
comparison, Pr0201 yields values between $0.38$ and $-0.76$~days -- the 
latter one is obtained when $J-K$ is used, and also becomes more negative 
if we use the higher frequency component as the rotational frequency. 
(Pr0201 has two closely spaced frequency components, corresponding 
to the periods of $5.63$ and $5.19$~days.)} 
Although Pr0211 is certainly not the only one with similar deviations, 
the values quoted correspond nearly $20$\% difference in rotational rate 
relative to `single normal' stars, which we consider interesting enough 
to consider a possible relation between the faster rotation and the 
planet-hosting status. 

If the cause of this difference is the angular momentum exchange 
between the planet and the host star, then from the models of \cite{brown2011} 
we expect this system to be near (within a few hundred million years) to the 
end of its life time. Unfortunately, the implied decrease of the orbital 
period would still be under a small fraction of a second over a year, 
so it is not easy to verify this hypothesis by direct observations. On the 
other hand, the fact that the two other planet hosts, Pr0201 in Praesepe and 
HD~285507 in the Hyades sit closer to the respective ridge lines is in 
agreement with the expectation that in the case of these longer period systems 
the tidal interaction becomes far less important, and change their rotational 
periods with the same pace as other, `normal' single stars. 

The migration of HJs and their interaction with their host stars is obviously 
a very complex problem. This complexity is highlighted by the recent result 
of \cite{mcquillan2013b}, who, based on the rotational analysis of $737$ 
planet host stars in the Kepler field, showed that fast rotation and short 
period planet hosting is a low probability event (see also 
\citealt{walkowicz2013} and for a possible theoretical explanation 
\citealt{teitler2014}). We note however, that there are several/many HJ 
host stars in the field that rotate fast. For example (just selecting a 
few representative cases), from the published $V_{\rm rot}\sin i$ and 
$R_{\rm star}$ values one can easily derive that WASP-33 and KELT-1 
are deeply within the `prohibited' lower left corner of the 
$P_{\rm orb}$ -- $P_{\rm rot}$ plot (see Fig.~2 of \citealt{mcquillan2013b}). 
The evolutionary ages are $0.4\pm0.30$ and $1.75\pm0.25$~Gyr, respectively. 
The age for WASP-33 is in agreement with the expected gyro age but for 
KELT-1 the gyro age is shorter by several factors. This indicates that KELT-1 
rotates faster than expected for a single star with the same physical parameters.   
Other hosts, such as WASP-103, HAT-P-32, HAT-P-41, among others, are definitely 
below the border line of the $P_{\rm orb}$ -- $P_{\rm rot}$ diagram, and, again, 
with multiple factors lower gyro ages as expected from the reasonably accurate 
evolutionary ages. These differences are large, even if we consider that there 
is a general inconsistency between the gyro and isochrone ages (e.g., 
\citealt{barnes2009}). The two hosts in Praesepe are at the border line, 
whereas the one in the Hyades is among the many other KOIs in the middle part 
of the diagram.

%
\section{Conclusions}
We analyzed $381$ high probability members of the galactic open cluster 
Praesepe and found that $47$\% of them show significant variability, best 
understood as a consequence of the cyclic variation of the spotted area 
of the stars due to rotation. We found a fairly tight color/magnitude -- 
period relation, in consonance with the earlier work of \cite{delorme2011}, 
based on a sample size less than a third of the one presented here. 
Our sample extends to the lower-amplitude variables at the hotter side, 
up to early F stars. This, and the dense coverage down to the late K regime, 
allows us to trace the nonlinearity in the color dependence of the rotational 
periods. Because of the high-rate sampling and wide color range covered 
by these data and because of the relatively accurate age of Praesepe, 
the derived color-period relation is well-suited for gyrochronological 
age determination (i.e., \citealt{skumanich1972} and \citealt{barnes2003}). 

We found a rather small number ($13$\%) of the rotational variables with 
coherent light variation; most of them ($39$\%) have residual power at the 
main rotational component and there are also many ($41$\%) that have harmonic 
components. A relatively small fraction ($7$\%) of them show separate but 
closely spaced frequency components, probably related to differential 
rotation. This latter type of variables are at the hotter, luminous end 
of the rotational sequence, whereas the ones with coherent single-period 
component are found preferably at the cooler (mid/low-luminosity) side. 
Each type of the other variables (the ones with unstable single component 
and those with harmonic components) cover uniformly the full color/magnitude 
range of rotational variables. The amplitudes show a broad range throughout 
the full span of color. An overall amplitude increase toward redder/fainter 
colors/magnitudes is also observable. However, this increase is largely 
attributed to the higher detection limits for those fainter stars. 

From the two planet hosting stars the cooler one (Pr0211) has a rotational 
period that is $\sim 2$~days shorter than the one predicted by the main 
rotational relation at the given color in this cluster. Although there 
are several other -- supposedly not planet host -- stars with similar 
deviations, we may interpret this shorter period as a possible result of 
the exchange of angular momentum between the star and the planet at the 
expense of the planet's orbital angular momentum. This inference is perhaps 
corroborated by the fact that the other planet (and also HD~285507 in the 
Hyades) are much closer to the corresponding ridge values. The planets in 
these systems have $2-3$-times longer orbital periods than the one in 
Pr0211, so we expect considerably weaker tidal interactions. Also, as 
discussed in Sect.~4, there are several Hot Jupiter planet hosts in the field 
that rotate faster than expected from their isochrone ages. The host of Pr0211~b 
could be one of these spun-up stars.

\section*{Acknowledgments}
We thank the referee Aleks Scholz for the valuable comments that helped to 
improve the paper. This publication makes use of data products from the Two 
Micron All Sky Survey, which is a joint project of the University of 
Massachusetts and the Infrared Processing and Analysis Center/California 
Institute of Technology, funded by the National Aeronautics and Space 
Administration and the National Science Foundation. This research was made 
possible through the use of the AAVSO Photometric All-Sky Survey (APASS), 
funded by the Robert Martin Ayers Sciences Fund, the SIMBAD database and the 
VizieR catalogue access tool, operated at CDS, Strasbourg, France. G.~K. thanks 
the Hungarian Scientific Research Foundation (OTKA) for support through grant 
K-81373. HATNet observations have been funded by NASA grants NNG04GN74G and 
NNX13AJ15G. G.\'A.~B., Z.~C. and K.~P. acknowledge partial support from NASA 
grant NNX09AB29G. J.~H. acknowledges partial support from NSF grant AST-1108686 
and NASA grant NNX14AF87G. K.P. acknowledges support from NASA grant NNX13AQ62G. 
S.~Q. acknowledges support from NSF grant DGE-1051030.

%
\appendix

%
\section{Non-rotational variables}
\begin{figure*}
\centering
  \begin{tabular}{ccccc}
    \includegraphics[width=.27\textwidth]{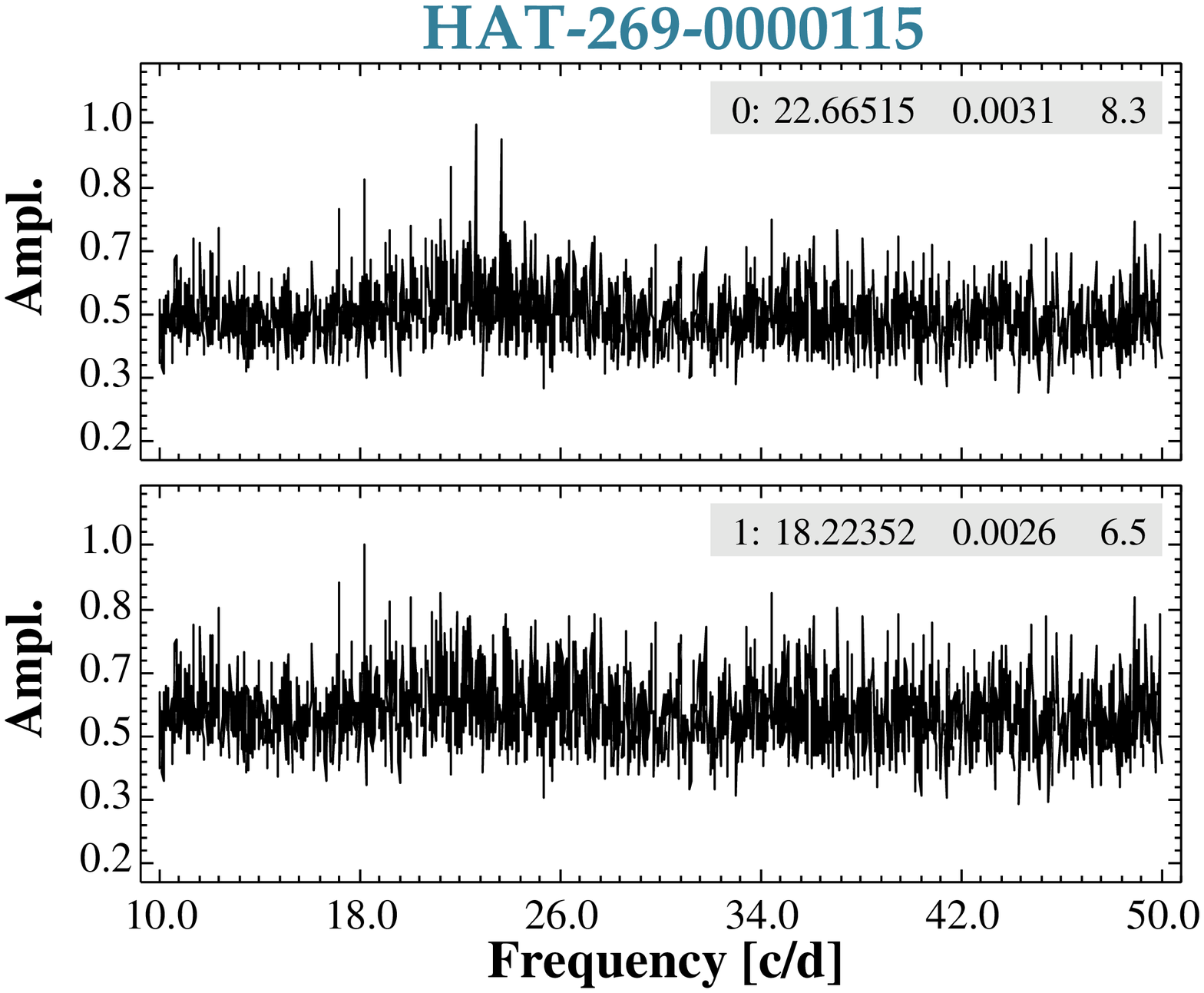} &
    \includegraphics[width=.27\textwidth]{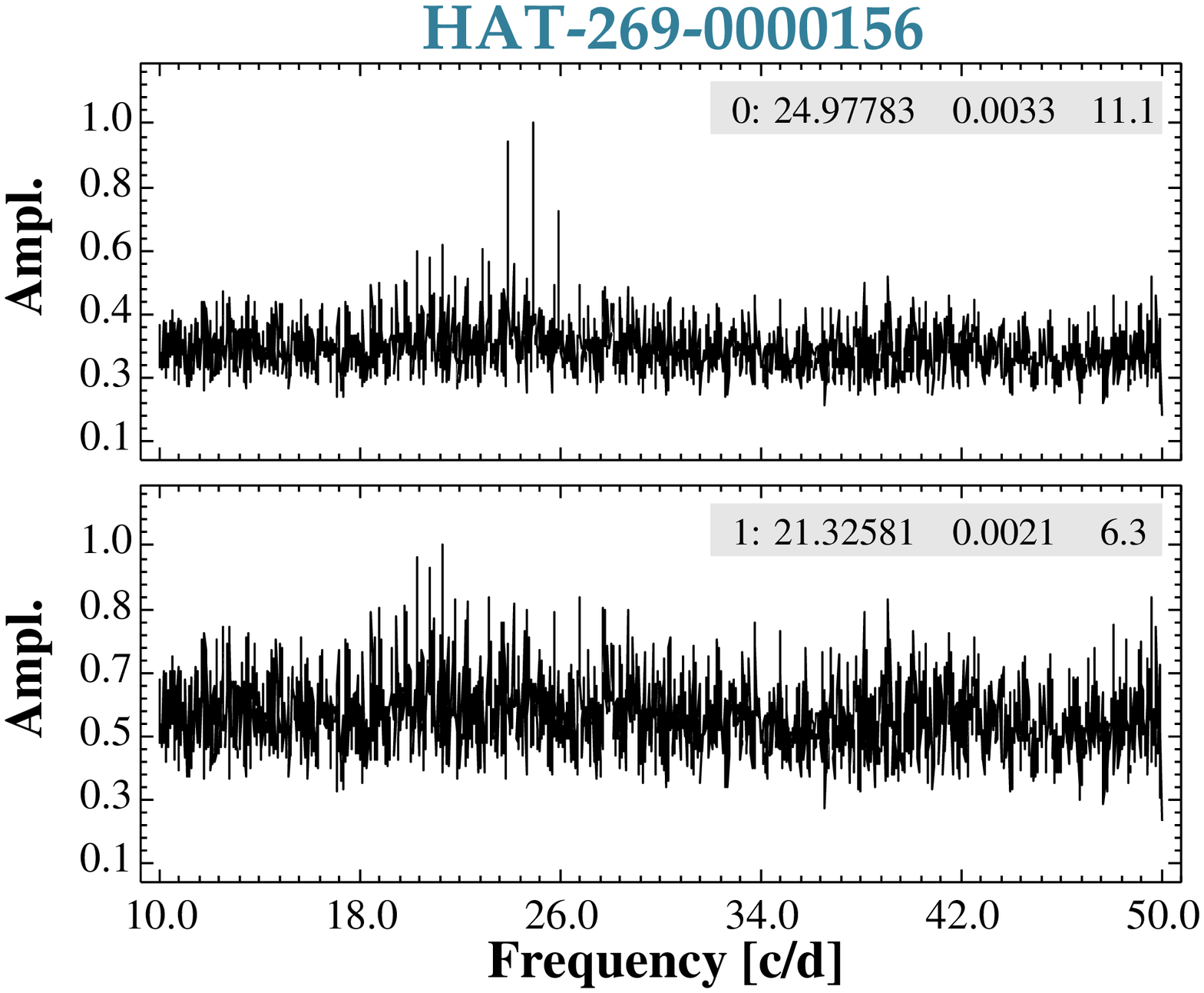} &
    \includegraphics[width=.27\textwidth]{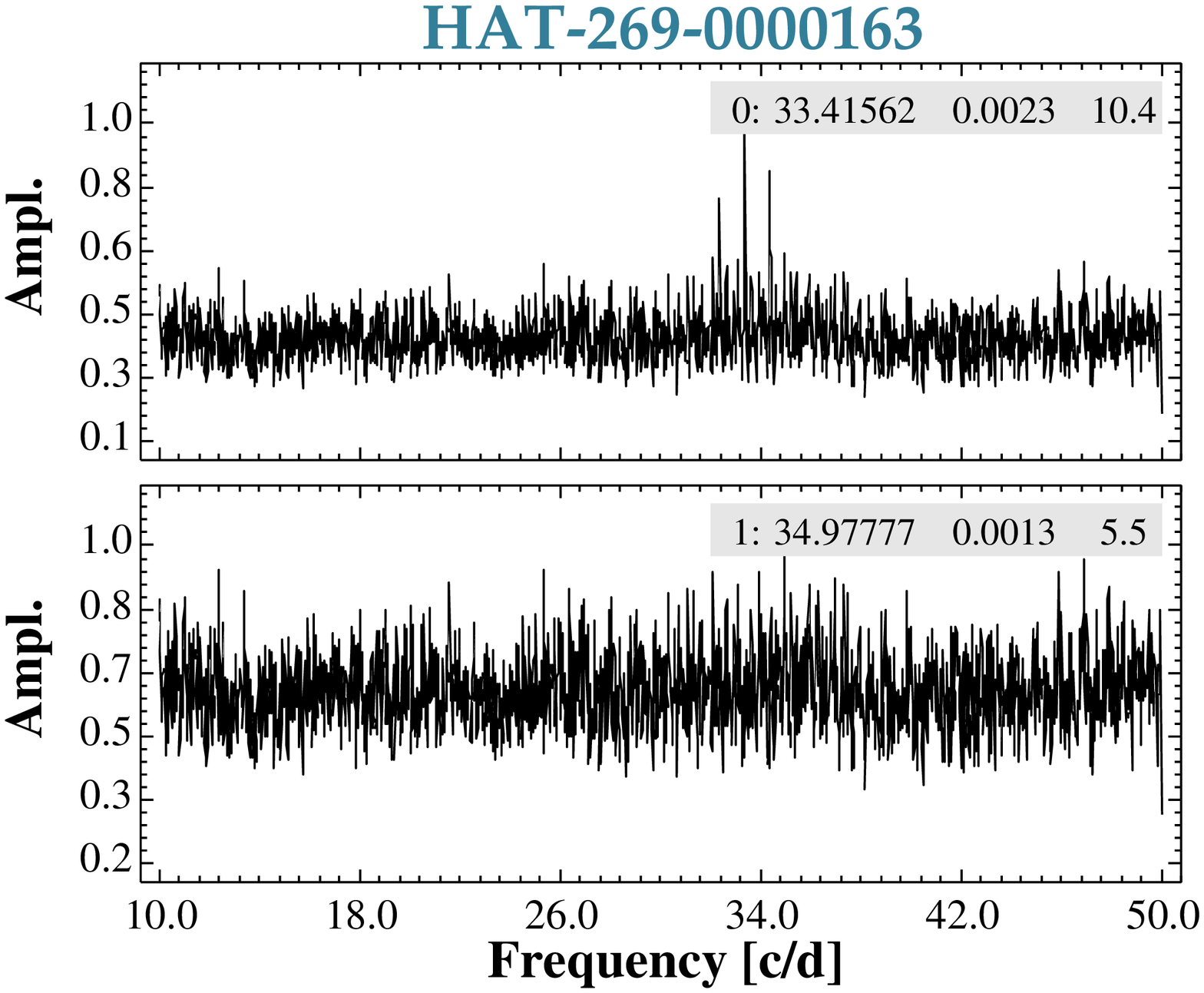} \\
    \includegraphics[width=.27\textwidth]{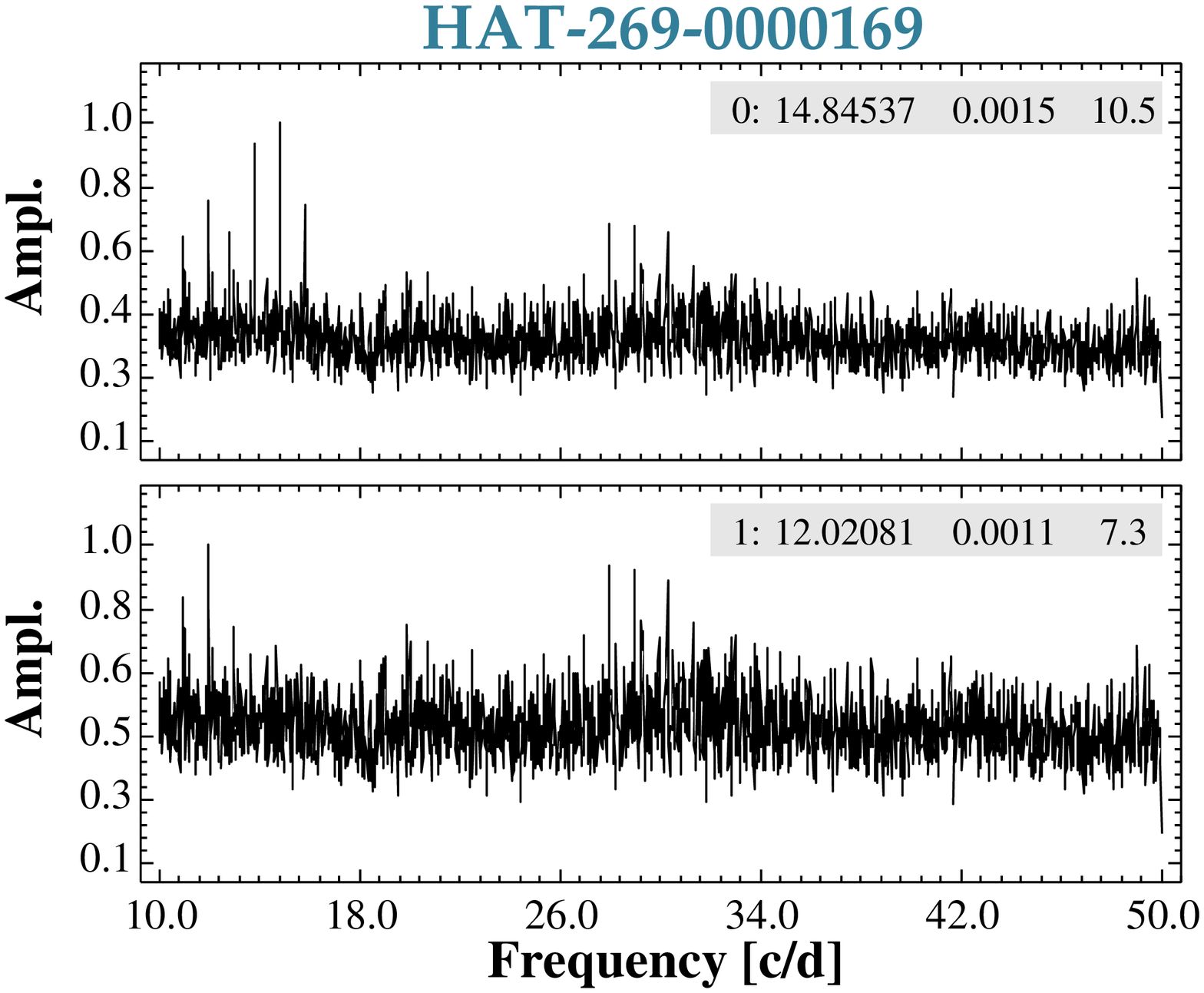} &
    \includegraphics[width=.27\textwidth]{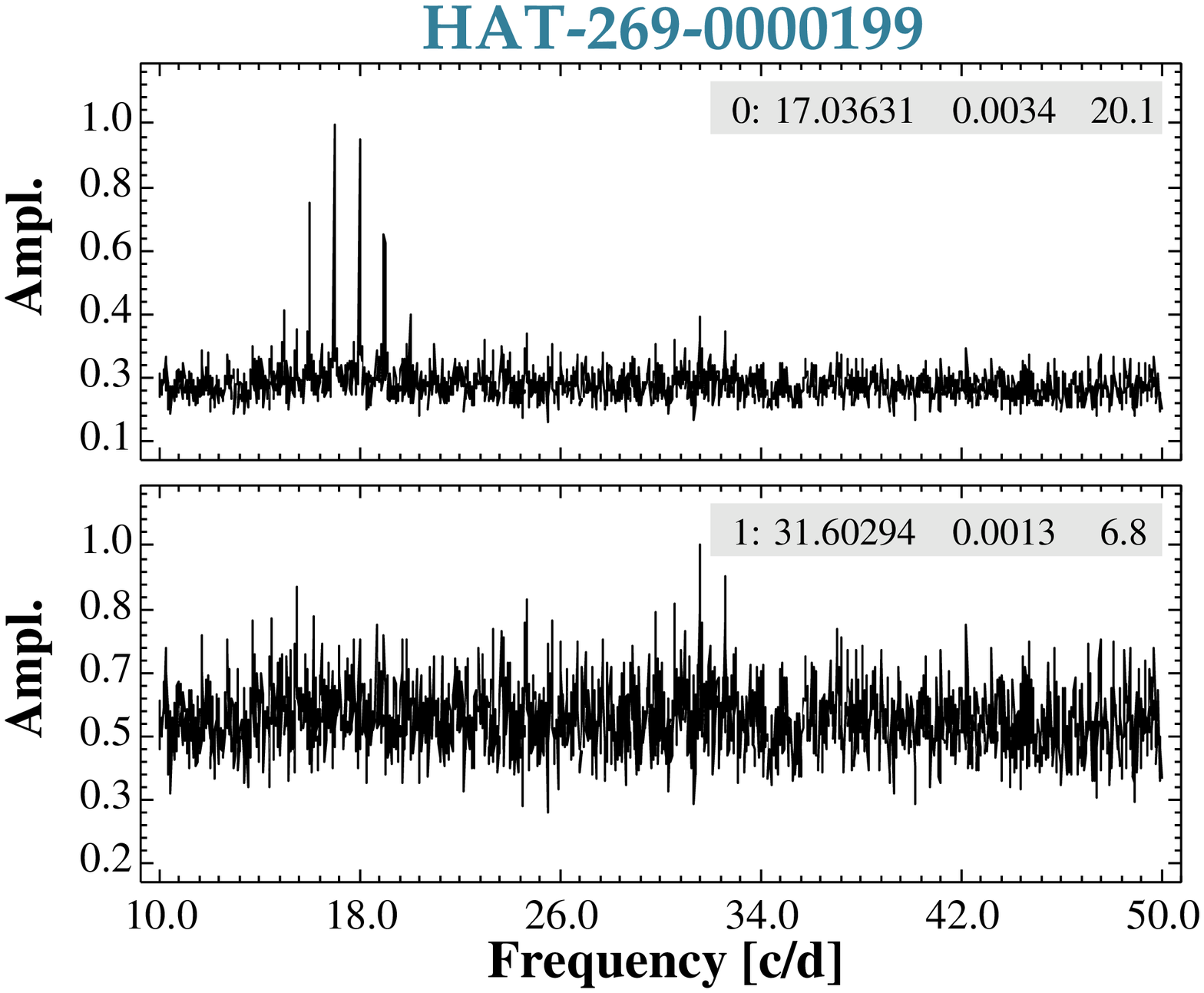} &
    \includegraphics[width=.27\textwidth]{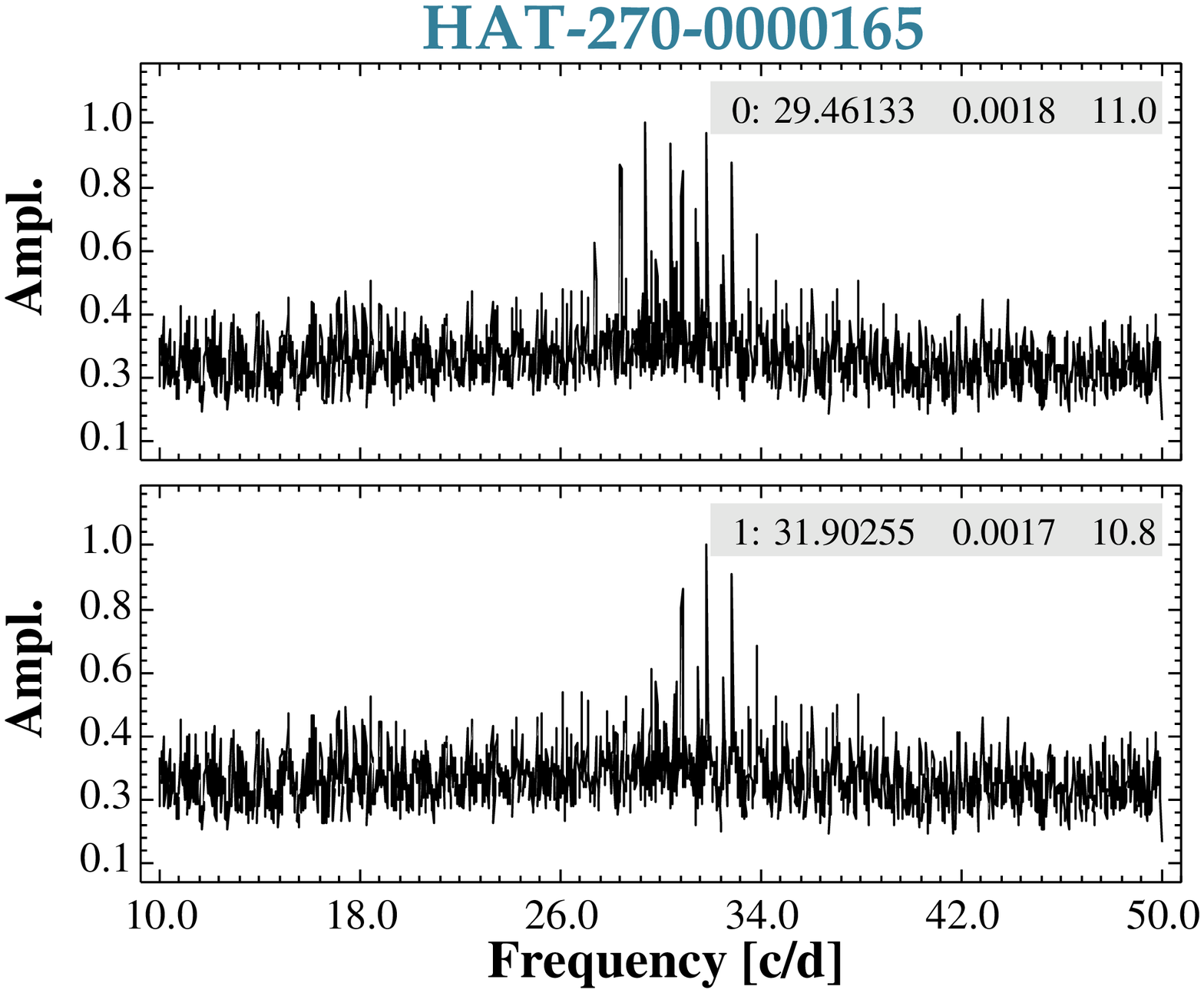} \\
    \includegraphics[width=.27\textwidth]{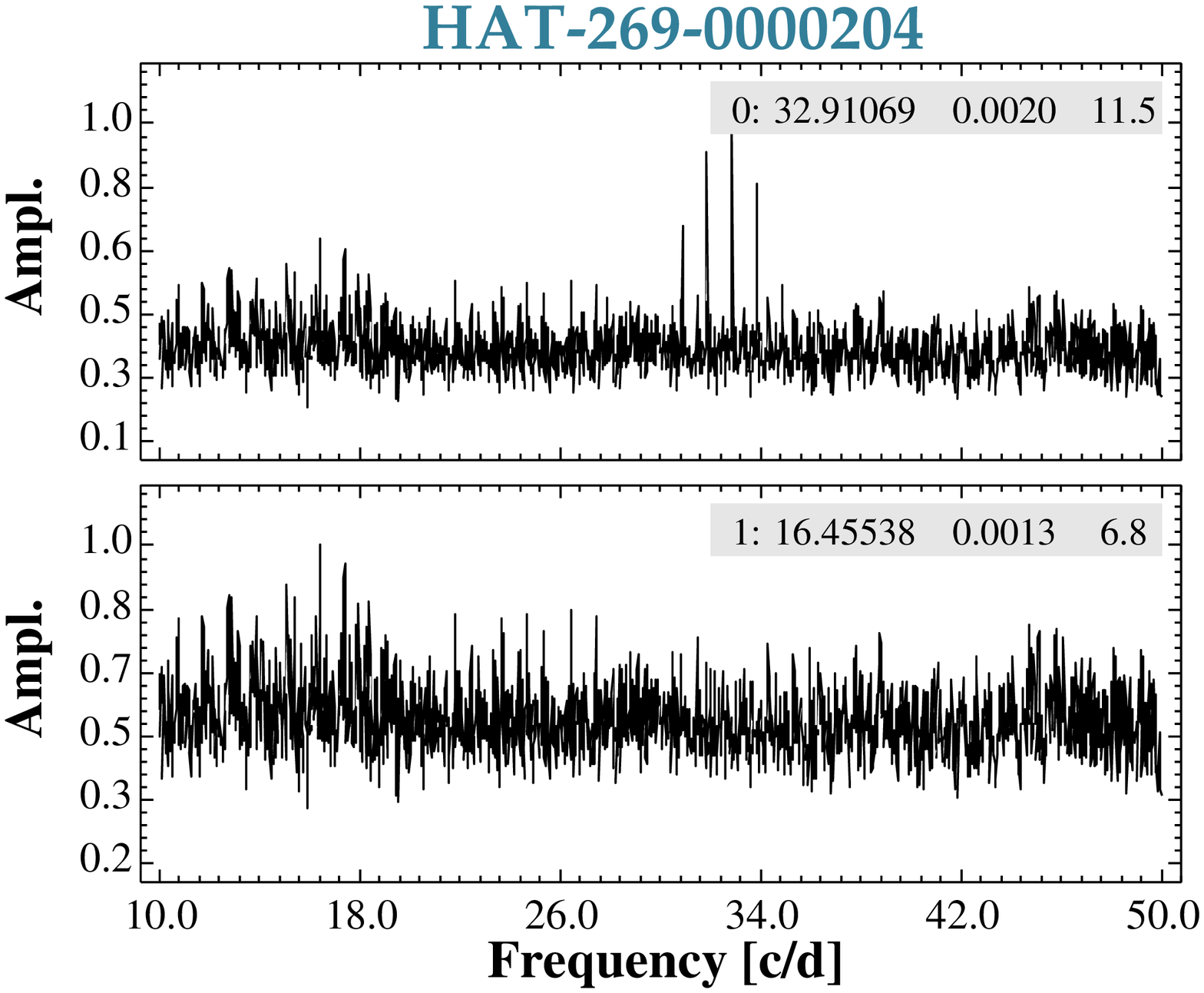} &
    \includegraphics[width=.27\textwidth]{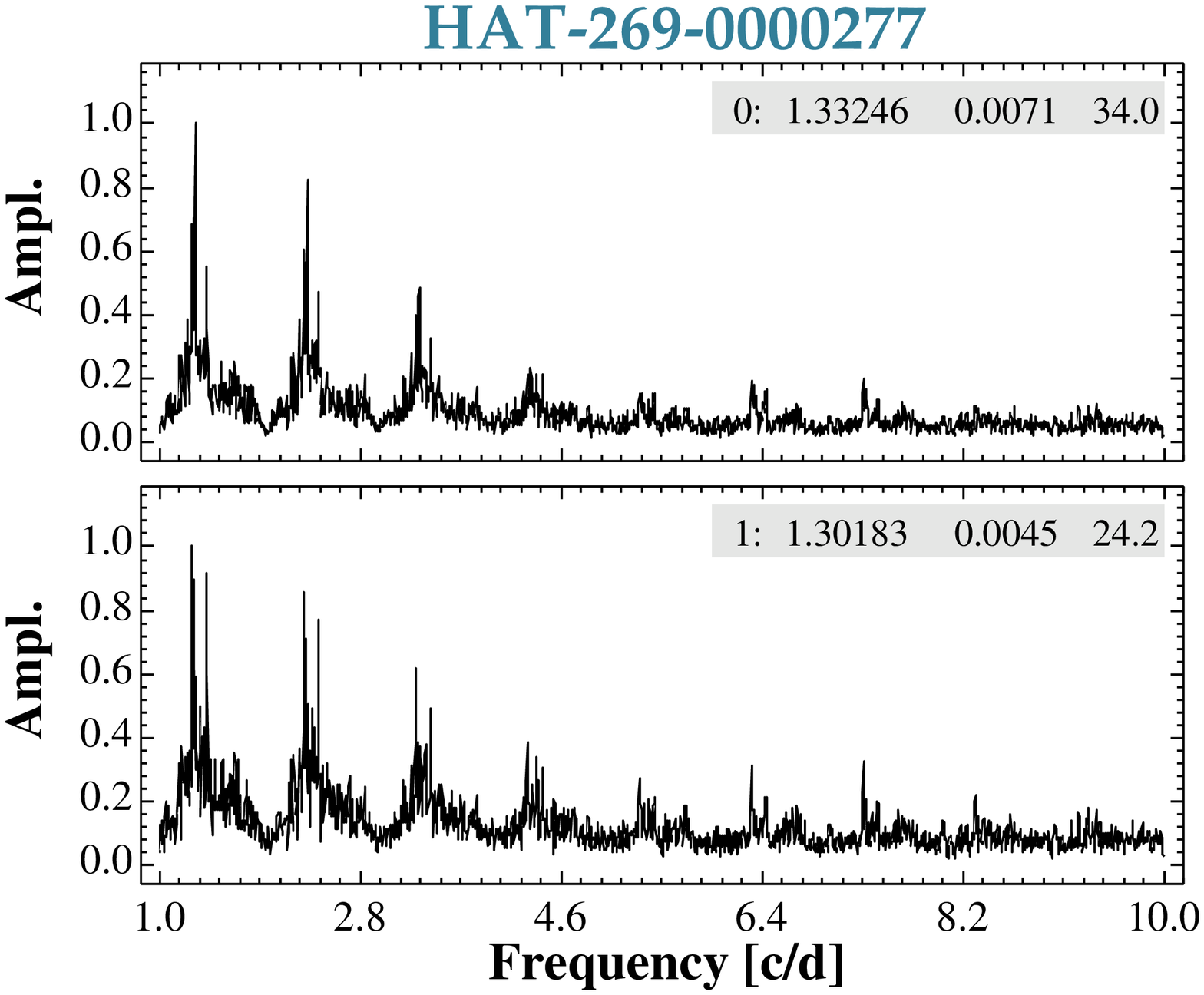} &
    \includegraphics[width=.27\textwidth]{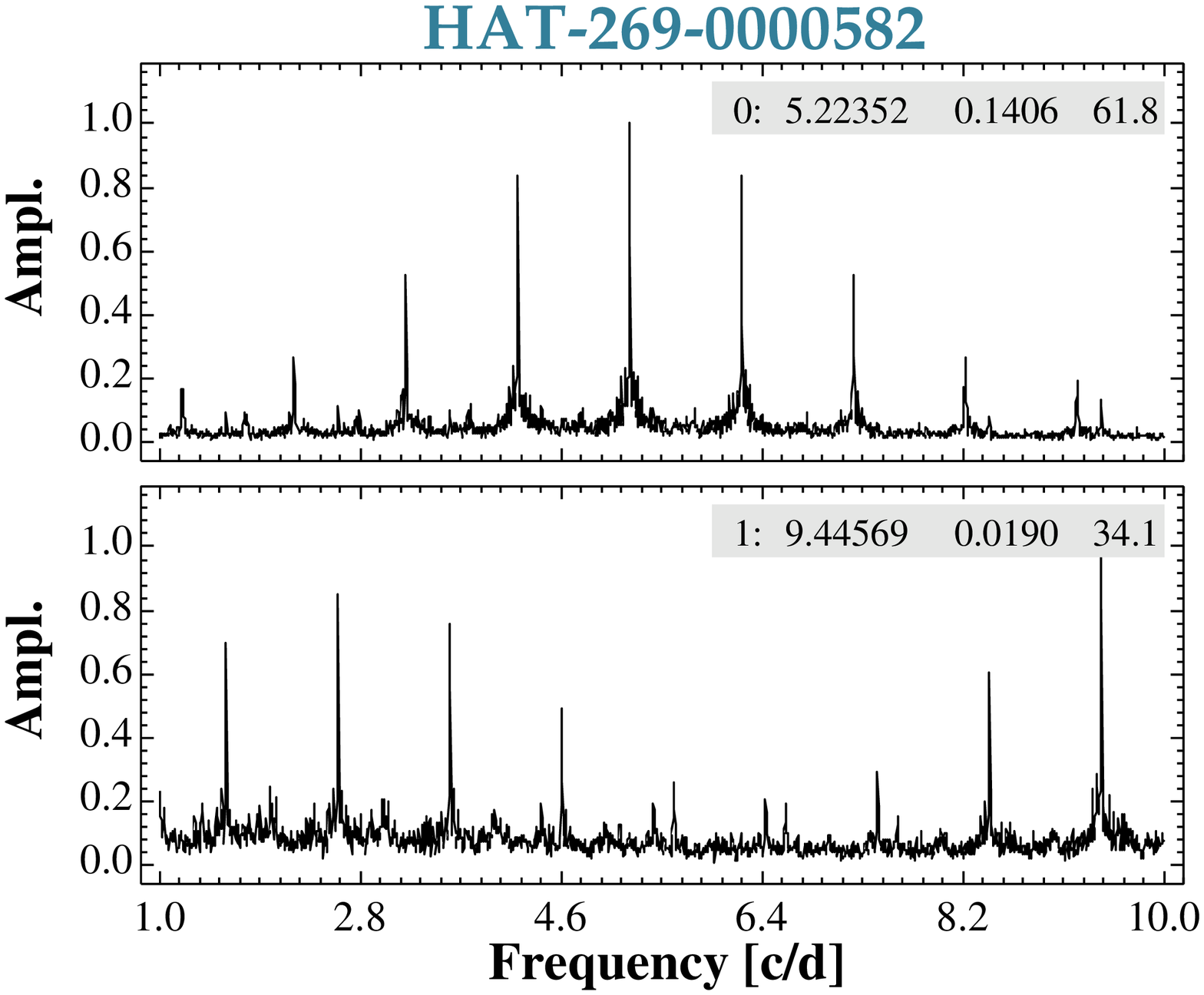} \\
  \end{tabular}
  \caption{Frequency spectra of the non-rotational variables with frequencies 
  higher than $1$~d$^{-1}$. In each sub-figure the upper and lower panels show, 
  respectively, the frequency spectra of the data and the spectra obtained 
  after the first pre-whitening (with the peak frequency shown on the upper 
  panel). The insets show the pre-whitening order, the peak frequency, the 
  amplitude and the SNR of the peak.}
\label{sp-nonrot}
\end{figure*}

\begin{figure}
 \vspace{0pt}
 \includegraphics[angle=0,width=85mm]{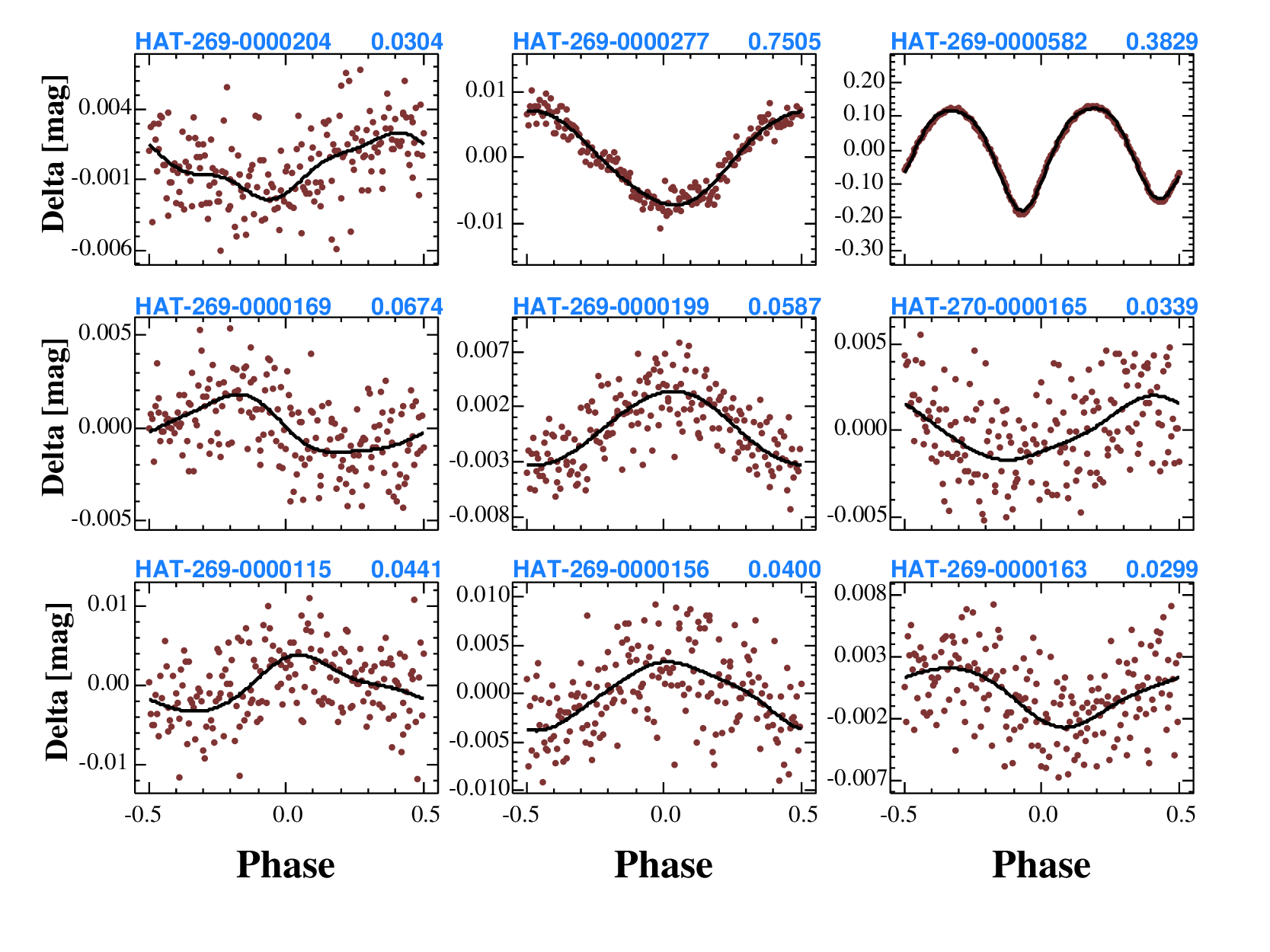}
 \caption{Folded light curves of the non-rotational variables found in 
 the analysis of the $381$ Praesepe members with periods shorter than 1~day. 
 For better visibility the folded data are binned in $200$ bins. The continuous 
 line is the 3-rd (for HAT-269-0000582 the 5-th) order Fourier fit. The 
 relatively large scatter in some cases is due to in part of the multiperiodic 
 nature of these, mostly $\delta$~Scuti-type variables. Except for the upper 
 two panels, the total amplitudes are smaller than those given in 
 Table~\ref{data-nonrot-var}, since we used the TFA-filtered light curves 
 without the reconstructive option.}
\label{lc-nonrot}
\end{figure}

By following the methodology employed in the search for rotational variables, 
we searched for other types of variables among the $381$ high-probability 
cluster members (see Sect.~2.1). We found $10$ variables in the frequency range 
of [0,50]d$^{-1}$. Table~\ref{data-nonrot-var} displays the most important 
properties of these variables. Figures~\ref{sp-nonrot} and \ref{lc-nonrot} 
show the frequency spectra and the folded light curves for variables with 
frequencies greater than $1$d$^{-1}$. The case of HAT-269-0000465 is special. 
It is KW~495, a triple spectroscopic system \citep{mermilliod1994}. Its frequency 
spectrum is rather complicated with two main frequency clumps at $\sim 0.079$ 
and $\sim 0.207$d$^{-1}$. The data are insufficient to classify this variable. 
Because of the detected period range, it is possible that the source of the 
signal is the rotation of two of the member stars. Alternatively, the 
complicated frequency spectrum and the late F (or early G) spectral type may 
suggest that this is a $\gamma$~Dor-type star.   

In a technical note we mention that in several cases the application of TFA 
filtering was necessary for an unambiguous detection. Since the time series 
are the result of the merging of the data from the individual fields, and 
our main interest is to detect variability, we do not employ signal 
reconstruction as the second phase in the TFA filtering process (see 
\citealt{kovacs2005}). As a result, except for HAT-269-0000582, where the signal
was stronger in the data before TFA filtering, the amplitudes are lower in 
the panels of Figures~\ref{sp-nonrot}, than the ones listed in 
Table~\ref{data-nonrot-var}, where we used the original (EPD) data to compute 
the total amplitudes. 

Most of the variables are known and classified as of $\delta$~Scuti type. 
We have probably 3 new discoveries (i.e., none of them are listed in the 
site of CDS nor mentioned in the current literature summary of this site 
on Praesepe). Star HAT-269-0000277 is most likely a pulsating variable 
with closely-spaced frequency components. This, together with the shape 
of the light curve, strongly suggests that this variable is a nonradial 
pulsator. The other object (HAT-270-0000165) is clearly a $\delta$~Scuti 
star. 

%
%

It is interesting to compare the frequencies we obtained with those 
of \cite{breger2012} derived from the data observed by the MOST satellite   
for the two variables common in these two studies. For BS Cnc \cite{breger2012} 
detected 20 frequencies. We have good agreement for the main component: 
$(\nu_0, {\rm Amp})=(17.0363, 0.0062)$ for MOST and $(17.03626, 0.0068)$ 
for HATNet (please note that our amplitude corresponds to the total signal 
amplitude - including also the [small] contribution of the other signal 
components). We also detected another component at $31.60299$~d$^{-1}$ with 
a much lower significance, at an amplitude nearly a third that of the main 
component. Interestingly, among the 20 components they found in the MOST 
data, there is the 10th component with basically the same frequency: 
$f_{10}=31.5988$. \cite{breger2012} mention the puzzling nature of this 
component, since it was actually absent in their 2008 dataset whereas had 
a relatively large amplitude of $2.1$~mmag in 2009. Our data cover the period 
of December 2008 and June 2011 (with a large gap between May 2009 and November 
2011). Therefore, it is likely that the amplitude of this component was still 
significant after 2009. 

The variability of HD~73872 was discovered by the MOST satellite. Some $18$ 
components were detected, with the main component at 
$(\nu_0, {\rm Amp})=(33.416, 0.0030)$ and a second one at $(35.981, 0.0025)$. We 
found a significant component at $33.41569$~d$^{-1}$ and a far less significant 
one as the second component at $34.97772$~d$^{-1}$. This latter one is very 
close to the $1$~d$^{-1}$ alias of the second component listed by 
\cite{breger2012}.   

%
%
\begin{table*}
 \begin{minipage}{200mm}
  \caption{Non-rotational variables in Praesepe detected in the HATNet 
  database}
  \label{data-nonrot-var}
  \begin{tabular}{cclrcccrl}
  \hline
   HAT ID & 2MASS ID & Other ID & $\nu$~[d$^{-1}$] & K~[mag] & J$-$K~[mag] &  A~[mag] & SNR & Type\\ 
  \hline
HAT-269-0000115 & 08420650+1924405 & BX Cnc   & 22.6652144 & 7.431 & 0.114 & 0.0072 &   8.3 & $\delta$~Scuti\\
HAT-269-0000156 & 08374070+1931063 & BR Cnc   & 24.9778239 & 7.662 & 0.117 & 0.0078 &  11.1 & $\delta$~Scuti \\
HAT-269-0000163 & 08411377+1955191 & HD 73872 & 33.4156874 & 7.774 & 0.101 & 0.0054 &  10.4 & $\delta$~Scuti \\
HAT-269-0000169 & 08405247+2015594 & BW Cnc   & 14.8453080 & 7.804 & 0.107 & 0.0036 &  10.5 & $\delta$~Scuti \\
HAT-269-0000199 & 08390909+1935327 & BS Cnc   & 17.0362610 & 7.875 & 0.131 & 0.0068 &  20.1 & $\delta$~Scuti \\
HAT-270-0000165 & 08452825+2023435 & HD 74587 & 29.4613325 & 7.864 & 0.148 & 0.0039 &  11.0 & $\delta$~Scuti \\
HAT-269-0000204 & 08403296+1911395 & BV Cnc   & 32.9106985 & 7.964 & 0.119 & 0.0047 &  11.5 & $\delta$~Scuti \\
HAT-269-0000277 & 08411067+1949465 & HD 73854 &  1.3324437 & 8.190 & 0.174 & 0.0148 &  34.0 & puls.? \\
HAT-269-0000465 & 08430593+1926152 & BD+19 2087 & 0.2071673& 8.461 & 0.333 & 0.0050 &  20.8 & misc.? \\
HAT-269-0000582 & 08400171+1859595 & TX Cnc   &  5.2235226 & 8.698 & 0.355 & 0.2983 &  61.8 & EB \\
\hline
\end{tabular}
\end{minipage}
\begin{flushleft}
\underline{Notes:}
For variables with peak frequency $1 < \nu < 10$~[d$^{-1}$] the signal-to-noise 
ratio (SNR) is computed in the $[1,10]$~d$^{-1}$ band. For those with 
$\nu > 10$~[d$^{-1}$], the $[10,50]$~d$^{-1}$ band is used. Amplitudes A~[mag] 
are peak-to-peak values from the 4$^{\rm th}$ order Fourier fit to the instrumental 
Sloan ``r'' light curves, without TFA filtering. For the special case of 
HAT-269-0000465 the amplitude is a rough visual estimate of the total range of 
variation. For other details of this object, see text.  
\end{flushleft}
\end{table*}
%

%

%
\begin{figure}
 \vspace{0pt}
 \includegraphics[angle=0,width=85mm]{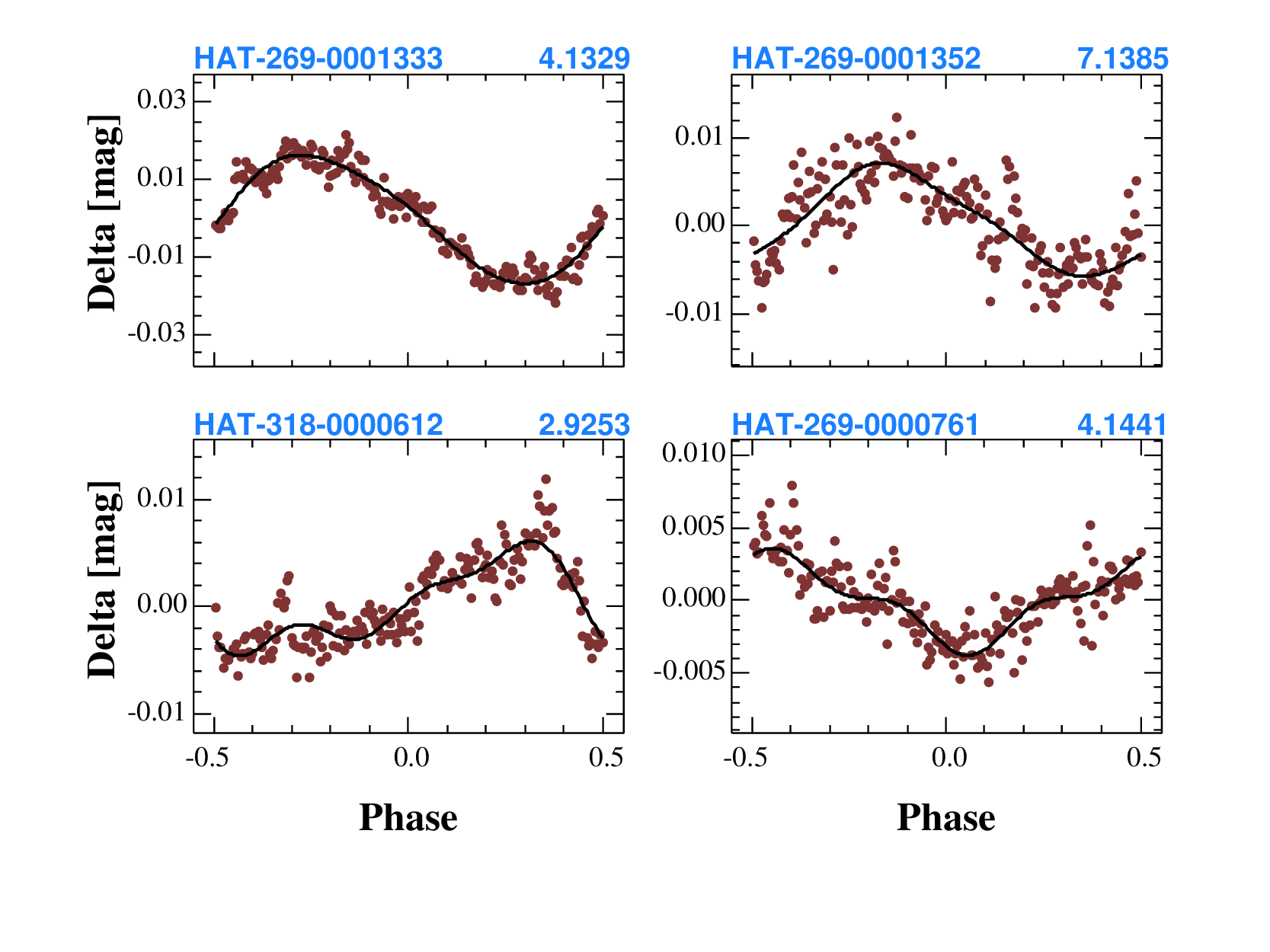}
 \caption{Folded light curves of the four rotational variables that have 
 close periods to the spectroscopic periods given by 
 \protect\cite{mermilliod2009}. For better visibility the data are binned 
 in $200$ bins. The continuous line is the 3-rd order Fourier fit. The 
 original (EPD) data are used, without TFA filtering.}
\label{sb-lc}
\end{figure}

%
%
\section{Spectroscopic binaries}
The status of the spectroscopic binaries (SBs) associated with the cluster 
has been summarized recently by \cite{mermilliod2009}. Based on their data 
collected by the CORAVEL spectrograph over a period of nearly two decades, 
they presented the orbital elements of 29 SBs. When we cross-correlate these 
stars with the $381$ stars selected from the overlap of the list of cluster 
members given by \cite{kraus2007} and the HATNet database, we get a list of 
$24$ stars. The spectroscopic (orbital) and photometric frequencies of these 
objects are shown in Table~\ref{sb}. The photometric frequency is the 
one that corresponds to the highest peak in the spectrum and does not 
necessarily agree with the finally adopted rotational frequency (see 
Table~\ref{data-rot-var} for comparison). As discussed in the Sect.~6.1, 
the variable type of HAT-269-0000465 is unclear in spite of its high SNR.    

We see that the orbital periods are, in general, much longer than the 
rotational periods. There are only three cases when we might face some 
ambiguity in distinguishing between the light variations caused by 
eclipses and by spots. However, inspecting the folded light curves does 
not lend too much support to the eclipse hypothesis, since they all seem 
to have distorted sinusoidal shape, with no resemblance of an eclipsing 
binary (see Figure~\ref{sb-lc}). There is a fourth variable (HAT-269-0000761), 
which is an outlier in the color-period diagram with orbital and rotational 
periods differing by some $30$\%. We could not find any trace of the orbital 
period in the photometric data. However, when we compare the photometric 
frequency with the one computed from the rotational velocity of 
$11.9$~kms$^{-1}$ \citep{mermilliod2009}, we get 
$f_{\rm rot}=0.2547$~d$^{-1}$ (see Sect.~3.1 for details of the estimation 
of the necessary physical parameters of the star). This is in a very nice 
agreement with our photometric frequency.  
 
%
%
\begin{table*}
 \begin{minipage}{200mm}
  \caption{Spectroscopic binaries in Praesepe}
  \label{sb}
  \scalebox{1.00}{
  \begin{tabular}{crrcrccr}
  \hline
HAT ID  &  KW ID  &  K  & J$-$K & f$_{\rm phot}$  & f$_{\rm orb}$  & f$_{\rm orb}$/f$_{\rm phot}$ & SNR\\
 \hline
HAT-269-0000913 &  508 &  7.293 &  0.300 & 0.16167 &0.00154 & 0.0095 & 22.9\\
HAT-269-0000528 &   47 &  7.353 &  0.338 & 6.28626 &0.02889 & 0.0046 &  5.2\\
HAT-269-0000761 &  181 &  7.431 &  0.360 & 0.24131 &0.17047 & 0.7064 & 23.1\\
HAT-269-0000489 &  268 &  7.610 &  0.259 & 0.58068 &0.00693 & 0.0119 &  5.4\\
HAT-269-0000396 &  416 &  7.637 &  0.217 & 1.37448 &0.03870 & 0.0282 &  5.0\\
HAT-269-0000726 & 3532 &  7.662 &  0.358 & 1.11693 &0.02944 & 0.0264 &  3.5\\
HAT-269-0000465 &  495 &  7.769 &  0.333 & 0.21501 &0.02786 & 0.1296 & 17.8\\
HAT-269-0000598 &  549 &  7.774 &  0.262 & 4.17972 &0.03196 & 0.0076 &  5.9\\
HAT-269-0001333 &  434 &  7.790 &  0.383 & 0.24196 &0.25428 & 1.0509 & 38.4\\
HAT-318-0000612 & 3655 &  7.804 &  0.422 & 0.34184 &0.33537 & 0.9811 & 23.7\\
HAT-269-0001352 &   55 &  7.864 &  0.535 & 0.14009 &0.16708 & 1.1927 & 22.0\\
HAT-269-0000767 &  287 &  7.875 &  0.363 & 1.45366 &0.00014 & 0.0001 &  4.8\\
HAT-269-0000556 &  365 &  7.944 &  0.359 & 0.12007 &0.01952 & 0.1626 &  7.2\\
HAT-269-0001570 &  368 &  7.946 &  0.430 & 0.11103 &0.01306 & 0.1176 & 10.9\\
HAT-269-0000850 &  325 &  7.964 &  0.396 & 0.32065 &0.00346 & 0.0108 &  8.5\\
HAT-269-0000287 &  142 &  8.044 &  0.241 & 0.37165 &0.02175 & 0.0585 &  4.9\\
HAT-269-0000372 &  439 &  8.051 &  0.207 & 8.88541 &0.00218 & 0.0002 &  4.8\\
HAT-269-0001402 &  539 &  8.063 &  0.396 & 0.26847 &0.00018 & 0.0007 & 20.8\\
HAT-269-0000968 &  127 &  8.100 &  0.329 & 0.90909 &0.07530 & 0.0828 &  6.2\\
HAT-269-0007340 &  536 &  8.190 &  0.873 & 0.58212 &0.00079 & 0.0014 &  5.2\\
HAT-269-0000794 &  556 &  8.221 &  0.366 & 0.22052 &0.00700 & 0.0317 & 26.2\\
HAT-269-0001490 &  184 &  8.351 &  0.502 & 0.09536 &0.02108 & 0.2211 & 13.0\\
HAT-317-0001780 & 2025 &  8.369 &  0.503 & 0.20720 &0.00038 & 0.0018 & 18.6\\
HAT-269-0003949 & 1184 &  8.377 &  0.660 & 9.57013 &0.81290 & 0.0849 &  3.7\\
\hline
\end{tabular}}
\end{minipage}
\begin{flushleft}
\underline{Notes:}
Photometric (f$_{\rm phot}$) and orbital (f$_{\rm orb}$) frequencies 
(in [d$^{-1}$]), respectively, are taken from this paper and from 
\cite{mermilliod2009}. The signal-to-noise ratio (SNR) refers to the peak 
frequency in the $[0,10]$d$^{-1}$ band, and is computed from the Fourier 
spectra of the TFA-filtered data. See text for more about f$_{\rm phot}$ 
and on the status of HAT-269-0000465. 
\end{flushleft}
\end{table*}
%

%

\end{document}